\documentclass[manuscript,nonacm]{acmart}
\renewcommand\footnotetextcopyrightpermission[1]{}
\usepackage{color, colortbl}
\usepackage{lscape} 
\usepackage{colortbl}  
\usepackage{tikz, pgfplots}
\usepackage{enumitem}
\usepackage{multirow}
\usepackage{multicol}
\usepackage{mdframed} 
\usepackage{adjustbox}
\usepackage{rotating, graphicx, wrapfig}
\usepackage{subcaption}
\usepackage{url} 
\usepackage{supertabular,booktabs}
\usepackage[graphicx]{realboxes}
\usepackage{subcaption} 
\usepackage{longtable}

\usepackage{listings}
\definecolor{lightgray}{gray}{0.9}

\mdfdefinestyle{RQFrame}{
 outerlinewidth=0pt,
 skipabove=0pt,
 skipbelow=0pt,
 innertopmargin=6pt,
 innerbottommargin=0pt,
 linewidth=0pt,
 topline=false,
 rightline=false,
 leftline=false,
 innerrightmargin=4pt,
 innerleftmargin=4pt}

\newcounter{RQCounter}
\newcommand{\RQ}[2]{
\refstepcounter{RQCounter} \label{#1}
\begin{mdframed}[style=RQFrame]\noindent
    \textbf{RQ}$_{\arabic{RQCounter}}$.~\emph{#2}
\end{mdframed}
}
\newcommand{\hr}[1]{\textbf{RQ}$_{\ref{#1}}$}
\definecolor{lightgray}{gray}{0.9}
\definecolor{Gray}{gray}{0.9}

\usepackage{hyperref} 

\usepackage{graphicx} 

\AtBeginDocument{%
  }

\begin{document}

% \newcounter{RQCounter}
% \newcommand{\hr}[1]{\textbf{RQ}$_{\ref{#1}}$}

%%
%% The "title" command has an optional parameter,
%% allowing the author to define a "short title" to be used in page headers.
\title{In-person, Online and Back Again - A Tale of Three Hybrid Hackathons}

%%
%% The "author" command and its associated commands are used to define
%% the authors and their affiliations.
%% Of note is the shared affiliation of the first two authors, and the
%% "authornote" and "authornotemark" commands
%% used to denote shared contribution to the research.
\author{Abasi-amefon Obot Affia-Jomants}
\affiliation{%
  \institution{University of Tartu}
  \city{Tartu}
  \country{Estonia}}
\orcid{0000-0001-7627-1198}
\email{amefon.affia@ut.ee}

\author{Alexander Serebrenik}
%\authornote{Both authors contributed equally to this research.}
\affiliation{%
  \institution{Eindhoven University of Technology}
  \city{Eindhoven}
  \country{The Netherlands}}
\orcid{0000-0002-1418-0095}
\email{a.serebrenik@tue.nl}

\author{James D. Herbsleb}
%\authornote{Both authors contributed equally to this research.}
\affiliation{%
  \institution{Carnegie Mellon University}
  \city{Pittsburgh}
  \country{USA}}
\orcid{0000-0002-7159-7524}
\email{a.serebrenik@tue.nl}

\author{Alexander Nolte}
%\authornote{Both authors contributed equally to this research.}
\affiliation{%
  \institution{Eindhoven University of Technology}
  \city{Eindhoven}
  \country{The Netherlands}}
\affiliation{%
  \institution{Carnegie Mellon University}
  \city{Pittsburgh}
  \country{USA}}
\orcid{0000-0003-1255-824X}
\email{a.u.nolte@tue.nl}
\renewcommand{\shortauthors}{Affia et al.}

%%
%% The abstract is a short summary of the work to be presented in the
%% article. 

\begin{abstract} 
Hybrid hackathons, which combine in-person and online participation, present unique challenges for organizers and participants.  
Although these events are increasingly conducted globally, research on them remains fragmented, with limited integration between hackathon studies and hybrid collaboration. 
Existing strategies for in-person or online-only events often fail to address the unique challenges of hybrid formats, such as managing communication across physical and virtual spaces and ensuring balanced participation. 
Our work addresses this gap by examining how hybrid hackathons function through the lens of hybrid collaboration theories, analyzing how organizers structure these events and how participants navigate hybrid-specific challenges.  
Drawing on established theories of hybrid collaboration, we examine key dimensions -- synchronicity, physical distribution, dynamic transitions, and technological infrastructure -- that shape collaboration in hybrid events. Through an exploratory case study of three hackathon events involving observations and interviews with organizers and participants, we analyze how these dimensions are implemented and their effects on participant experiences.
Our findings reveal differing organizer considerations of the hybrid dimensions in the hackathon design, leading to distinct experiences for participants. Implementation styles -- favoring in-person, online, or balanced participation -- led to varied participant experiences, affecting access to resources, communication, and team coordination. Organizers in our study also often relied on technology to bridge hybrid interactions, but sometimes overlooked critical aspects like time-zone management, dynamic transitions, and targeted support for hybrid teams. Additionally, participants in their teams responded to gaps in event scaffolding by adapting collaboration strategies, underscoring that hybrid formats are still not fully integrated into hackathon planning and revealing gaps in organizers’ preparedness for hybrid events. 
Learning from our findings, we offer practical recommendations when organizing hybrid hackathon events and recommendations to participants when attending hybrid hackathon events.  
 
\end{abstract}

%%
%% The code below is generated by the tool at http://dl.acm.org/ccs.cfm.
%% Please copy and paste the code instead of the example below.
%%
%\begin{CCSXML}
%<ccs2012>
%  <concept>
%    <concept_id>10003120.10003121.10011748</concept_id>
%    <concept_desc>Human-centered computing~Empirical studies in HCI</concept_desc>
%    <concept_significance>500</concept_significance>
%  </concept>
%  <concept>
%    <concept_id>10003120.10003130.10003131.10003235</concept_id>
%    <concept_desc>Human-centered computing~Collaborative content creation</concept_desc>
%    <concept_significance>500</concept_significance>
%  </concept>
%  <concept>
%    <concept_id>10010147.10010257</concept_id>
%    <concept_desc>Computing methodologies~Machine learning</concept_desc>
%    <concept_significance>300</concept_significance>
%  </concept>
%</ccs2012>
%\end{CCSXML}

%\ccsdesc[500]{Human-centered computing~Empirical studies in HCI}
%\ccsdesc[500]{Human-centered computing~Collaborative content creation}
%\ccsdesc[300]{Computing methodologies~Machine learning}

%%
%% Keywords. The author(s) should pick words that accurately describe
%% the work being presented. Separate the keywords with commas.
%% A "teaser" image appears between the author and affiliation
%% information and the body of the document, and typically spans the
%% page.

%\received{20 February 2007}
%\received[revised]{12 March 2009}
%\received[accepted]{5 June 2009}

%%
%% This command processes the author and affiliation and title
%% information and builds the first part of the formatted document.
\maketitle

\section{Introduction} \label{sec:introduction}

Hackathons are a global phenomenon. They are time-bounded events, attracting participants with diverse backgrounds and expertise to work on a shared team project to create an artefact~\cite{falk2024future}. 
Hackathons have grown popular because they create a unique environment that fosters rapid innovation and creativity~\cite{heller2023hack}, provide experiential learning opportunities~\cite{schulten2024we,affia2022integrating}, and bring together diverse participants where individuals can connect with potential employers, collaborators, and mentors~\cite{warner2017hack}.

Hackathons have been commonly studied as in-person events, but during the global pandemic of 2020 and 2021, it became clear that online events can provide valuable collaboration spaces as well~\cite{mendes2022socio}. Moreover, they expanded collaboration opportunities~\cite{yokoi2021crisis}, allowing individuals to participate who otherwise would not have had the chance due to financial, visa-related, scheduling, or other issues~\cite{wang2022learnings,falk2024future,mendes2022socio}. It was thus reasonable that after the pandemic, when there was a strong drive to go back to in-person events, hackathon organizers would start to experiment with hybrid setups that blend physical and virtual participation. 
Prior work on in-person~\cite{pe2019understanding} and online hackathon events~\cite{wang2022learnings,mendes2022socio} is valuable; still, hybrid events introduce distinct challenges. These include managing communication across physical and virtual spaces, ensuring equitable participation for both co-located and remote participants, and addressing the technical complexities of integrating multiple tools and platforms~\cite{neumayr2021hybrid}.  
Strategies developed for either in-person or online-only hackathon events overlook these hybrid-specific challenges, making them insufficient for fully addressing the demands of hybrid formats.   
While hybrid hackathons are being conducted globally, research on them remains limited. Existing studies have explored hybrid hackathons in relation to goals such as education and skill building~\cite{gama2023comfort,jonsson2022digital}, teamwork~\cite{porras2021experiences}, and industry skills~\cite{skrupskaya2022international}, but have not systematically examined how these events are designed and experienced in light of broader theories of hybrid collaboration. Our study addresses this gap by examining how organizers prepare for and run hybrid hackathons and how the event scaffolding can impact participants and their teams.

\RQ{RQ1}{How do organizers prepare for and run hybrid hackathons?} 

Since organizers already run hybrid hackathons, understanding how they plan and run them is essential to addressing the research gap, as it reveals the real-world strategies used to manage the hybrid format in their respective events.  Although models of hybrid collaboration exist~\cite{neumayr2021hybrid}, it is unclear whether and how organizers take them into account when organizing their events. It is also unclear whether these approaches are sufficient since they have been developed in environments where collaboration is more controlled, such as workplace collaborations or long-term projects~\cite{neumayr2021hybrid,Wu2023Information,bula2024nurturing,deschenes2024digital}. Hackathons, by contrast, are fast-paced events~\cite{than20182nd,falk2024future} where small, diverse teams work intensively over a short period, often without prior interaction or familiarity~\cite{than20182nd}. They present useful sites for studying hybrid work due to short-term goals and ad-hoc team formation.
    
\RQ{RQ2}{How do the hybrid hackathons' scaffolding impact participants and their teams?} 

Organizer considerations of the hybrid format and the eventual hackathon design influence what happens during the hackathon~\cite{chau2023hackathons} and form the hackathon scaffolding. Participants who attend the hackathon engage in this hackathon scaffolding, thus impacting their experience (\hyperref[RQ2]{\hr{RQ2}})~\cite{nolte2020organize,chau2023hackathons}. Since hackathons are, by their nature, collaborative events~\cite{falk2024future}, the impact of an event’s scaffolding will manifest itself in the way that teams collaborating across different modalities (remote, co-located, and hybrid) are formed, the way they collaborate, and the outcome of their collaboration~\cite{lyonnet2021hackathon}. 

To answer our research questions, we conducted an exploratory case study~\cite{runeson2012case} of three hybrid hackathon events. An exploratory case study is an appropriate research method for addressing our research questions because it allows for an in-depth investigation of organizer considerations and participant experiences in hybrid hackathons within their real-life context. 
The hybrid hackathon events were selected to represent a diversity of typologies~\cite{drouhard2016typology}, sizes~\cite{nolte2020organize}, duration~\cite{nolte2020organize}, audience~\cite{medina2020we}, and goal/themes~\cite{nolte2020organize}. Our selection criteria ensure that the cases reflect common but varied hybrid hackathon formats, enabling us to explore overarching patterns and specific, context-dependent challenges or advantages. 
Our case study approach allows for the collection of rich qualitative data through interviews and event observations. We interviewed organizers before and after their event to understand how they planned it and how it eventually turned out (\hyperref[RQ1]{\hr{RQ1}}). In addition, we observed teams during the event and conducted interviews with them afterward to study their experience participating in a hybrid hackathon (\hyperref[RQ2]{\hr{RQ2}}). 

Our findings indicate that, although one might expect fundamental hybrid dimensions such as synchronicity (managing synchronous and asynchronous interactions) and physical distribution (integrating remote and in-person spaces) to be consistently addressed by organizers, this was not always the case. Logistical concerns such as the size, goals, and audience of each event influenced whether or not organizers fully considered the hybrid format in their design. This often led to hackathon designs favoring one modality (in-person or remote) at the expense of the other, creating disparities in engagement, resource access, and the quality of team collaboration.
During the events, organizers in our study tended to rely on technology to manage interactions and integrate physical and virtual spaces. Some often overlooked aspects include accommodating time-zone differences, managing dynamic transitions between synchronous and asynchronous work, and targeting support for hybrid teams.
We also found that participants in our study responded to perceived gaps in the event scaffolding by adapting their collaboration strategies, often deviating from the organizer's expectations. These adaptations were influenced by the need for convenience in collaboration, flexibility in participation, and access to support, showing how teams compensated for the shortcomings in the hybrid design.
These variations and adaptations, where they impacted hybrid collaboration, indicated that the hybrid format is not yet fully understood or integrated into hackathon planning. This reveals gaps in knowledge or resources available to organizers when transitioning from fully in-person or fully online formats to hybrid ones.

The contribution of this paper is twofold. First, we contribute to the understanding of hybrid collaboration in hackathons by examining the rationale behind organizers' decisions in structuring their hybrid events and exploring the implications of these choices on participant engagement and collaboration. Second, it offers practical recommendations for future organizers aiming to enhance the execution of hybrid hackathons and for participants seeking to maximize their experience and effectiveness in hybrid settings.
The remainder of this paper is organized as follows: Section~\ref{sec:background} reviews models of hybrid collaboration and their hybrid dimensions as a framework for analyzing prior work in hybrid hackathons. Section~\ref{sec:empirical-method} describes the study’s exploratory case study approach, participant recruitment, and data collection methods used to investigate hybrid hackathons. Section~\ref{sec:findings-hybrid-dimension} presents findings across hybrid dimensions, illustrating how organizer scaffolding of the hackathons influenced participant experiences and collaboration dynamics. Section~\ref{sec:discussion} examines implications for theory and practice, offering practical recommendations for organizers to enhance hybrid hackathon experiences and for participants to maximize their engagement and coordination.

\section{Background} \label{sec:background}

In Section~\ref{sec:what-hybrid}, we introduce and define the broader concept of \emph{hybrid collaboration}, drawing from existing dimension-based models~\cite{augstein2022towards}.  
In Section~\ref{sec:hybrid-hack}, we focus on how hybrid collaboration is implemented in hackathon settings, providing a foundation for understanding hybrid hackathons.

\subsection{Hybrid Collaboration}\label{sec:what-hybrid}
The term ``hybrid'' has been used to describe collaborative work environments combining both physical and virtual modes~\cite{neumayr2021hybrid,nussli2024creating}. This setting allows two or more co-located participants to collaborate with one or more remote participants, leveraging technological tools to bridge the spatial divide and facilitate interaction~\cite{olson2000distance,duckert2023collocated}. 
Hybrid settings also often reproduce asymmetries between co-located and remote participants -- such as unequal access to shared artifacts, informal exchanges, or peripheral awareness -- which can limit the visibility and participation of remote members~\cite{bjorn2024achieving}. These systemic asymmetries are often navigated tactically by participants, who adapt to constraints not fully addressed in the strategic design of hybrid events~\cite{bodker2016happenstance}. The proliferation of hybrid environments further challenges participants to coordinate presence and participation across multiple sites~\cite{buyukguzel2023spatial}, manipulate distributed objects~\cite{lok2004toward}, and operate within overlapping digital and physical modes~\cite{jordan2009blurring}. As a result, hybrid collaboration demands new approaches to interaction, task ownership, and resource management~\cite{nussli2024creating,bjorn2024achieving}.
Hybrid collaboration (termed mixed-presence collaboration in the past~\cite{robinson2007distributed}) has also become increasingly popularized due to the globalization of workforces and the flexibility required in collaborative work during the COVID-19 pandemic~\cite{neumayr2019hybrid,neumayr2022territoriality,avdullahu2023requirements}. 
A traditional example of the hybrid collaboration model is Johansen's time-space model~\cite{johansen1988groupware}, which categorizes groupware along two dimensions -- time (``same time'' vs. ``different time'') and space (``same place'' vs. ``different place''). These dichotomies are used to categorize collaborative activities into four distinct quadrants -- same time, same place (e.g., in-person meetings), same time, different place (e.g., phone or video calls), different time, same place (e.g., using a shared physical document in a meeting room), different time, different place (e.g., asynchronous email exchanges)~\cite{johansen1988groupware}. While this dichotomous approach was useful for understanding groupware and collaborative systems at the time it was developed~\cite{johansen1988groupware}, hybrid collaborative environments often involve fluid transitions between synchronous and asynchronous work and between co-located and remote interactions, blurring the boundaries between these categories~\cite{neumayr2018domino}. As a result, models proposed by Neumayr~\cite{neumayr2018domino} and Lee and Paine~\cite{lee2015matrix} demonstrate that hybrid collaboration emerges when the continua of both time and space intersect, making it distinct from purely co-located or purely remote collaboration models~\cite{neumayr2019hybrid}.   
Specifically, the Model of Coordinated Action (MoCA) by Lee and Paine~\cite{lee2015matrix} presents \textit{\textbf{synchronicity and asynchronicity}} -- blending synchronous and asynchronous interactions -- and \textit{\textbf{physical distribution}} -- blending physical and virtual spaces. Neumayr \textit{et al.}~\cite{neumayr2018domino} describe the interplay between these dimensions as \textit{\textbf{dynamic transitions}} between time and space, capturing the fluid movement across these continua, which defines the adaptive and flexible nature of hybrid collaboration.
The \textit{\textbf{technological infrastructure}} (typically comprising multiple tools and platforms) plays a critical role in facilitating these transitions, enabling teams to move seamlessly between synchronous and asynchronous interactions and between physical and virtual spaces~\cite{neumayr2021hybrid,neumayr2018domino}. Tools such as Slack\footnote{\url{https://slack.com/signin}}, GitHub\footnote{\url{https://github.com/}}, and Zoom\footnote{\url{https://zoom.us/join}} allow hybrid teams to collaborate across the full spectrum of Johansen’s model and support these dynamic shifts.
Lastly, we consider the complexities of collaborative groups in hybrid environments. Hybrid collaboration involves two or more co-located participants collaborating with one or more remote participants~\cite{neumayr2021hybrid,olson2000distance} working across different locations and interacting at varying times, where multiple coupling styles can coexist. Thus, we consider the \textit{\textbf{team size and dynamic}} as impacting hybrid collaboration.

We use these dimensions from hybrid collaborative models as a framework to analyze hybrid hackathons~\cite{augstein2022towards}. This approach aligns with how organizers may typically approach hackathon design~\cite{augstein2022towards}. 
In Section~\ref{sec:hybrid-hack}, we discuss the relevance of the dimensions of hybrid collaboration -- \textit{synchronicity and asynchronicity}, \textit{physical distribution}, \textit{dynamic transitions}, \textit{technological infrastructure}, and \textit{team size and dynamics} -- to the hackathon setting.

\subsection{Hybrid Hackathons}\label{sec:hybrid-hack}

Much of the existing research on hackathons has focused on fully in-person or fully online formats~\cite{falk2024future}. In-person hackathons benefit from real-time, face-to-face interactions, fostering creativity and rapid problem-solving~\cite{trainer2016hackathon}, while online hackathons depend on digital tools to facilitate collaboration across distributed teams~\cite{mendes2022socio}. However, hybrid hackathons -- that blend elements from both in-person and online formats -- introduce a new set of challenges, necessitating unique strategies for collaboration, team management, and technological integration.
The hybrid format has gained traction due to its flexibility~\cite{gama2023comfort,skrupskaya2022international}, adaptability~\cite{falk2024future}, and inclusivity~\cite{porras2021experiences}, particularly following the COVID-19 pandemic~\cite{falk2024future}. The format requires robust synchronous and asynchronous tools for effective collaboration~\cite{falk2024future}. However, existing studies have mainly focused on specific outcomes, such as education~\cite{gama2023comfort,jonsson2022digital}, teamwork~\cite{porras2021experiences}, and industry skills~\cite{skrupskaya2022international,khan2021innovation}. These studies often do not address how the hybrid collaboration dimensions -- synchronicity and asynchronicity, physical distribution, technology use, and team dynamics -- shape the overall participant and organizer experience. These dimensions are also critical to understanding the differences in ``hybrid'' hackathons compared to fully in-person or online ones.

\begin{enumerate}
    \item \textit{Synchronicity and Asynchronicity}: Communication is key in the early stages of collaboration and hybrid work~\cite{bula2024nurturing}; so, the absence of ongoing conversation due to integrating the remote setting can lead to misaligned ideas~\cite{mendes2022socio}. While in-person hackathons rely almost entirely on synchronous, real-time interactions~\cite{trainer2016hackathon}, and online hackathons may incorporate synchronous and asynchronous tools~\cite{mendes2022socio}, hybrid hackathons necessitate deliberate coordination of both synchronous and asynchronous modes of communication~\cite{mendes2022socio}.  
    Current studies, such as those by Porras et al.\cite{porras2021experiences} and Khan et al.\cite{khan2021innovation}, have mainly focused on synchronous activities, often starting with in-person interactions and transitioning to online collaboration. Gama et al.~\cite{gama2023comfort} also mention using technologies such as Slack or Discord to foster synchronous and asynchronous communication, but they do not address how these support collaboration within hybrid teams. 

    \item \textit{Physical Distribution}: The physical distribution of participants -- where some are co-located in-person, and others participate remotely -- and their spatial arrangements directly impact how participants engage with the event.  Porras et al.\cite{porras2021experiences} and Gama et al.\cite{gama2023comfort} note that physical perks such as meals or co-working spaces encouraged in-person participation. Also, the idea of maintaining local and online organizers to manage logistics across locations is touched upon by Gama et al.~\cite{gama2023comfort}. Still, there is little exploration of how integrating physical and virtual spaces affects hybrid teams during the events.

    \item \textit{Dynamic Transitions}: In-person hackathons take advantage of physical proximity for spontaneous and synchronous collaboration, while online hackathons facilitate distributed collaboration through digital platforms that allow for synchronous and asynchronous communication. Existing research, such as Porras et al.~\cite{porras2021experiences} and Skrupskaya et al.~\cite{skrupskaya2022international}, acknowledges the importance of in-person meetings to build team dynamics and foster team spirit. However, the studies do not follow participant engagement as collaboration shifts modality, how these transitions affect team productivity, particularly in events spanning multiple days (e.g., maintaining momentum after an overnight break), and how this impacts hackathon results. 

   \item \textit{Technological Infrastructure}: Unlike in online hackathons, where technology is the primary mode of interaction~\cite{schulten2022participants,mendes2022socio}, in hybrid events, technology must be integrated into both physical and virtual spaces. Studies like Porras et al.~\cite{porras2021experiences}, Skrupskaya et al.~\cite{skrupskaya2022international}, and Gama et al.~\cite{gama2023comfort} mention the use of platforms like Zoom and Slack in addition to the role of asynchronous tools, such as Google Drive and GitHub~\cite{porras2021experiences}. However, they do not explore how these technologies are incorporated into hybrid event infrastructure, the challenges posed by those technologies in hybrid hackathons, and organizer and participant adaptations~\cite{bodker2016happenstance}. Studies also don't emphasize how these tools might fail to integrate both modalities~\cite{bjorn2024achieving}, thus creating disjointed experiences for participants who must navigate multiple technologies across modalities.
   
   \item \textit{Team Size and Dynamics}: Research outside of hackathons has shown that team size impacts participation, with smaller teams fostering more active engagement~\cite{bradner2003effects,bradner2005team}. In hackathons, smaller, cross-functional teams of around four members from diverse backgrounds are found to be more effective, while larger teams can improve coordination by dividing tasks into subgroups~\cite{raittila2022forming}. However, the complexities of managing team size in hybrid hackathons -- whether small or large -- remain unexplored in the studied literature. Hybrid settings, where teams are split between physical and virtual spaces, introduce additional challenges in leadership, coordination, and resource allocation~\cite{grzegorczyk2021blending} that do not exist in fully in-person or fully online settings. While some studies on hybrid hackathons, such as Skrupskaya et al.\cite{skrupskaya2022international} and Khan et al.\cite{khan2021innovation} highlight the benefits of in-person meetings for team cohesion, they do not fully examine how hybrid teams function once collaboration progresses.  
\end{enumerate}
Our paper fills this gap by examining multiple hybrid hackathons across diverse contexts through the lens of these hybrid dimensions.
By analyzing how organizers prepare for and run hybrid hackathons and how participants experience the scaffolding that supports them, we provide actionable suggestions and best practices for organizers. These suggestions also extend to how participants can better prepare for hybrid hackathons.

\section{Empirical Method} \label{sec:empirical-method}
To answer our research questions, we performed an exploratory multiple case study research~\cite{runeson2012case,wohlin2021case} of three hybrid hackathon events. 
Drawing from Wohlin~\cite{wohlin2021case} and Runeson~\textit{et al.}~\cite{runeson2012case}, a case study is: \textit{an empirical inquiry that investigates a contemporary phenomenon within its real-life context using multiple sources of evidence, mainly when the boundary between the phenomenon and context is not evident, without the investigators actively influencing the outcomes of the case.}
This approach is appropriate because it allows us to study a new phenomenon -- organizer considerations of the hybrid format in event organization (\hyperref[RQ1]{\hr{RQ1}}) and participant experience of this hackathon scaffolding (\hyperref[RQ2]{\hr{RQ2}}) --  that has not been observable before in the context it occurs to draw insights about ``how'' it takes place~\cite{runeson2012case,wohlin2021case}.
Our study is framed by the hybrid dimensions (introduced in Section~\ref{sec:what-hybrid}) and organizer considerations of these dimensions in their hackathon setting (\hyperref[RQ1]{\hr{RQ1}}). We also focus on teams as the unit of analysis as we explore participants' experiences of the hackathon scaffolding -- in our case, the design of the event a team participates in --  (\hyperref[RQ2]{\hr{RQ2}}).
Our study has received approval from the Institutional Review Board (IRB), ensuring that all data collection procedures adhere to ethical guidelines for research involving human subjects. 
The following sections provide an overview of our study setting, participant recruitment, data collection, and analysis procedure.

\subsection{Study Setting} \label{sec:studysetting} 

Hackathons come in different sizes and forms and are organized in different domains for different goals. We conducted a multi-case study to get a comprehensive understanding of hybrid hackathons that extend beyond a single event.  
We adopted the hackathon typology outlined by Drouhard, Tanweer, and Fiore~\cite{drouhard2016typology} because it offers a framework that categorizes hackathons based on their primary \textit{goals} into three categories: (i) \textit{Communal hackathons} focus on building resources, infrastructure, or culture within a community, with the primary goal of fostering professional development and community building. (ii) \textit{Contributive hackathons} aim to contribute to a larger ongoing project by breaking the work into modular tasks that participants work on individually or in teams. (iii) \textit{Catalytic hackathons} showcase the utility of a dataset, technology, or approach, challenging participants to generate new, innovative ideas.
Using this framework, we further categorized hybrid hackathons by \textit{size}~\cite{medina2020we}, \textit{duration}~\cite{medina2020we}, \textit{audience}~\cite{nolte2020organize}, and \textit{theme}~\cite{nolte2020organize}.  

Thus, we selected three hybrid hackathon events that were open to public participation (both in-person and online), representing distinct hackathon categories (c.f. Table~\ref{tab:hackeventdetails}). Although these events are recurring and typically held annually, our study focused on a single instance of each hackathon. These events varied in size, audience, disciplinary domain, and goals, allowing us to capture a broad range of hybrid hackathon practices. Each hackathon was independently organized, with no overlap in organizers or participants, and the participant compositions reflect the distinct nature of each event. While the cases are diverse, they reflect typical forms of hybrid hackathons found in practice, ensuring relevance to the broader study of hybrid collaboration.
Our goal was to gain in-depth insights into how teams collaborate within the scaffolding of hybrid hackathons. Rather than aiming for maximum coverage, we purposefully selected a small number of teams across the three events to capture diverse perspectives based on team modality, size, and communication approaches. We view this sample as sufficient for the aims of our study. While the findings are not generalizable beyond these contexts, they offer rich insight into how collaboration unfolded across different hybrid hackathon configurations. 

\begin{table}[htp!]
\centering
\caption{Study setting and hackathon selection}
\vspace{-5pt}
\resizebox{\textwidth}{!}{%
\begin{tabular}% 
{|p{0.18\linewidth}|p{0.25\linewidth}|p{0.28\linewidth}|p{0.25\linewidth}|}
\hline
 & H1          & H2    & H3                        \\ \hline 
 \textbf{Event Date}   & October 2023 & November 2023  & June 2024 \\ \hline
 \textbf{Typology}~\cite{drouhard2016typology}    & Communal & Catalytic  & Contributive \\ \hline
\textbf{Size} (approx)~\cite{nolte2018you}    & 603 participants & 12 participants  & 55 participants \\ \hline
\textbf{Duration}~\cite{nolte2018you}  & 24 hours (no break) & 2 days (overnight break) & 5 days (overnight break) \\ \hline
\textbf{Goal/theme}~\cite{nolte2018you}  &  Opportunity for students to build technology projects  & Spread interest and knowledge in a specialised open-source database & Additions to the a community project    \\ \hline
\textbf{Audience}~\cite{medina2020we}  & Students & Students/Researchers  & Practitioners/Researchers \\ \hline 
\end{tabular}}
\label{tab:hackeventdetails}
\vspace{-10pt}
\end{table}

\subsubsection{Hackathon 1 (H1)} 

H1 was a large collegiate event with about 603 participants, including in-person and virtual participants in attendance, hosted to provide students with a platform to build technology projects. The in-person part of the hackathon took place on the campus of a large North American University. The communal nature of this hackathon aligns with developing resources and promoting professional development within a student community.
The event occurred over 24 hours and spanned 2 days. It was the first in-person hybrid hackathon, where the primary emphasis was placed on the physical space, playing a central role in the event’s organization, with an opportunity for participants to join virtually. The hackathon concluded with presentations to judges and a non-competitive showcase.

\subsubsection{Hackathon 2 (H2)}
H2 was a three-day (48-hour) event to spread knowledge and interest in an open-source database among students and researchers worldwide. The in-person part of the hackathon took place on the campus of a large North American University. This hackathon required participants to demonstrate the utility of a resource or approach, fitting the catalytic model, where participants focused on creating novel contributions.
The hackathon facilitated a hybrid format, with 12 (in-person and online) participants in attendance. The hackathon concluded with presentations to the other participants and feedback from mentors.

\subsubsection{Hackathon 3 (H3)}
H3 was a five-day event that was organized to expand and improve an existing handbook for reproducible, ethical, and collaborative data science. The in-person part of the hackathon took place at a research institute in the UK. The contributive model applies here, as participants worked on discrete tasks contributing to the larger, ongoing project (the handbook).
The event targets both new and existing contributors to the community resource, focusing on those who are familiar with the project but welcoming new contributors to the opportunity to join preparatory events to better understand the project and how they can contribute. The event promoted collaborative contributions, and participants could choose short ``contribution sessions'' to work collaboratively on the project's content. The H3 hackathon agenda comprised five of these contribution sessions throughout each day, spaced out from 08:00 UTC to 21:30 UTC. The hackathon concluded with final presentations shared with other participants.

\subsection{Participant recruitment}
We targeted both organizers and participants in each selected hybrid hackathon event to answer our research questions (\hyperref[RQ1]{\hr{RQ1}}, \hyperref[RQ2]{\hr{RQ2}}).  Organizers provide insights into the planning, setup, and execution of hybrid hackathons (\hyperref[RQ1]{\hr{RQ1}}), while participants offer accounts of their experience working in their teams within the hackathon scaffolding (\hyperref[RQ2]{\hr{RQ2}}).

For recruiting organizers, we employed two strategies. First, during the initial hackathon selection process, we identified key organizer contacts involved in the event's planning. These contacts were invited to participate in the study. Second, we asked these primary contacts to recruit additional organizers, specifically those in strategic roles, such as event coordination, technology management, and/or team facilitation. This ensured we gathered a more comprehensive picture of each hackathon organization from various organizers.
We also employed two strategies to recruit participants. First, we collaborated with the event organizers. The organizers supported us in sharing information about the research with hackathon participants, encouraging participants across hackathon modalities to sign up to participate. Second, by observation and after approval from the organizers, we approached and recruited participant teams, targeting participants seen to be actively involved in team collaborations during the event and particularly seeking teams across modalities.

\subsection{Data collection}\label{Sec:datacollection}

We collected qualitative data from event observations and participant and organizer interviews~\cite{affia2025research} to understand how organizers prepare for and run hybrid hackathons (\hyperref[RQ1]{\hr{RQ1}}) and how the scaffolding they provide impacts participants and their teams (\hyperref[RQ2]{\hr{RQ2}}). This approach aligns with the case study method by comparing and corroborating findings across various data sources and gaining complementary insights into findings.  As our interest was in how participants and organizers navigated hybrid-specific challenges during the event, we focused on qualitative methods rather than measuring productivity or output.
In this section, we introduce the study participants and how data collection was handled using our research instruments~\cite{affia2025research} to address our research questions.

\subsubsection{Observation}
During each hackathon, we conducted live, non-participant observations to understand how the hybrid format shaped participant collaboration, communication, and engagement within teams. Our role was strictly observational; we did not interact directly with participants or organizers during the event.
To guide the observation process, we developed a protocol focused on participant interaction, technology usage, team roles, task coordination, and spatial dynamics. Teams were selected for observation based on their initial modality (in-person, hybrid, or online), and we also noted organizer interventions—such as announcements or mentoring check-ins—when they directly influenced team activity. We documented observations using point-in-time photographs and written field notes, without using specialized tracking devices. These materials captured informal interactions and hybrid coordination practices in situ and were later used to:  \textit{(i)} contrast interview statements with observed behavior and;  \textit{(ii)} inform follow-up questions during interviews, allowing us to probe deeper into specific team practices and organizer decisions as they occurred during the event.
These observations provided contextual information about participant and organizer behaviors~\cite{stickdorn2018service}, which do not necessarily come out through the interviews.

\subsubsection{Interviews}

\paragraph{Organizers}

Organizers played a key role in setting up the hackathon. The purpose of the organizer interviews was to understand how organizers planned, structured, and managed the hybrid hackathons. 
We conducted semi-structured pre- and post-hackathon interviews to collect perceptions of how organizers considered these dimensions in their hackathon scaffolding, where participants in their teams collaborate at the hackathon event, thus addressing~\hyperref[RQ1]{\hr{RQ1}}.
We selected organizers based on their involvement in key aspects of hackathon planning and delivery to capture a range of perspectives on how hybrid events were scaffolded. In total, we interviewed 8 organizers across the three hackathons. In H1, we interviewed 3 organizers: two were responsible for participant onboarding and coordination, and one focused on technical support and platform setup. In H2, we interviewed 2 organizers: one technical coordinator and one main event organizer involved in major decision-making. In H3, we interviewed 3 organizers: one focused on operational delivery, one planning committee member supporting later time zones (Americas), and one leading contribution sessions in the local time zone. 

The 45-minute pre-hackathon interview consisted of three main parts. We started by first asking the organizers about their experience in planning hackathons and their motivations for organizing a hybrid hackathon (e.g. \textit{``How many hybrid hackathons have you organized before this one?'', ``How did you decide on a hybrid format instead of a purely in-person or online hackathon?''}). 
We also asked them about how they had planned for the hackathon in line with the hybrid dimensions (e.g., \textit{``What -- if anything -- did you do to prepare for the hybrid format? Prompt: [Participant communication], [Participant location (modality)], [Technology use], [Mentoring], [Participant collaboration]''}~\cite{neumayr2021hybrid,johansen1988groupware}). 
To target \textbf{synchronicity}, we asked questions about the organizers' plans for supporting both real-time and asynchronous communication and collaboration during the hackathon \textit{(e.g., ``How were you planning to support hybrid teamwork?'', ``How did mentor onboarding work? Were mentors made accessible to both online and in-person participants?'')}. For \textbf{physical distribution}, we focused on how organizers integrated physical and virtual spaces to support collaboration (e.g.,~\textit{``Can you describe the arrangements you made in the physical space to support the hybrid format?''}, and \textit{``From where did you expect remote participants to join?'')}. To address \textbf{technology infrastructure}, we asked about the specific tools used and anticipated challenges \textit{(e.g.,  ``What technologies were you using to support the hybrid format?'', ``What technology challenges did you anticipate and prepare for?'')}. Finally, to explore \textbf{team dynamics and structure}, we inquired about role assignments within hybrid teams and examples of how hybrid setups worked \textit{(e.g.,  ``How did you expect team formation for the hackathon? [spontaneous, pre-arranged, ...] and how did you support them?'', ``Were roles like moderator/leader suggested to the groups?'')}.
Lastly, we asked about any challenges faced while planning and expected during the hackathon event (e.g., \textit{``What were the challenges faced in planning [hackathon]?'', ``Can you share what challenges you might expect with the hybrid format?''}).

The 30-minute post-hackathon interview focused on reflections of the impact of the hackathon scaffolding implemented by the organizers (e.g., \textit{``Which - if any - issues did you become aware of related to the hybrid format during the hackathon?'', ``Which teams did you think the hybrid set-up worked very well for? Why was this the case?'', ``Any instances of teams you thought did not manage the hybrid format well? Can you describe what went wrong?''}).
These questions served as a basis for us to contextualize the organizers' responses and assess whether they made hybrid considerations for their hackathon scaffolding and their perceived impacts on the participants in their teams.

\begin{figure}[ht!]
%\vspace{-10pt}
	\centering\includegraphics[width=1\textwidth]{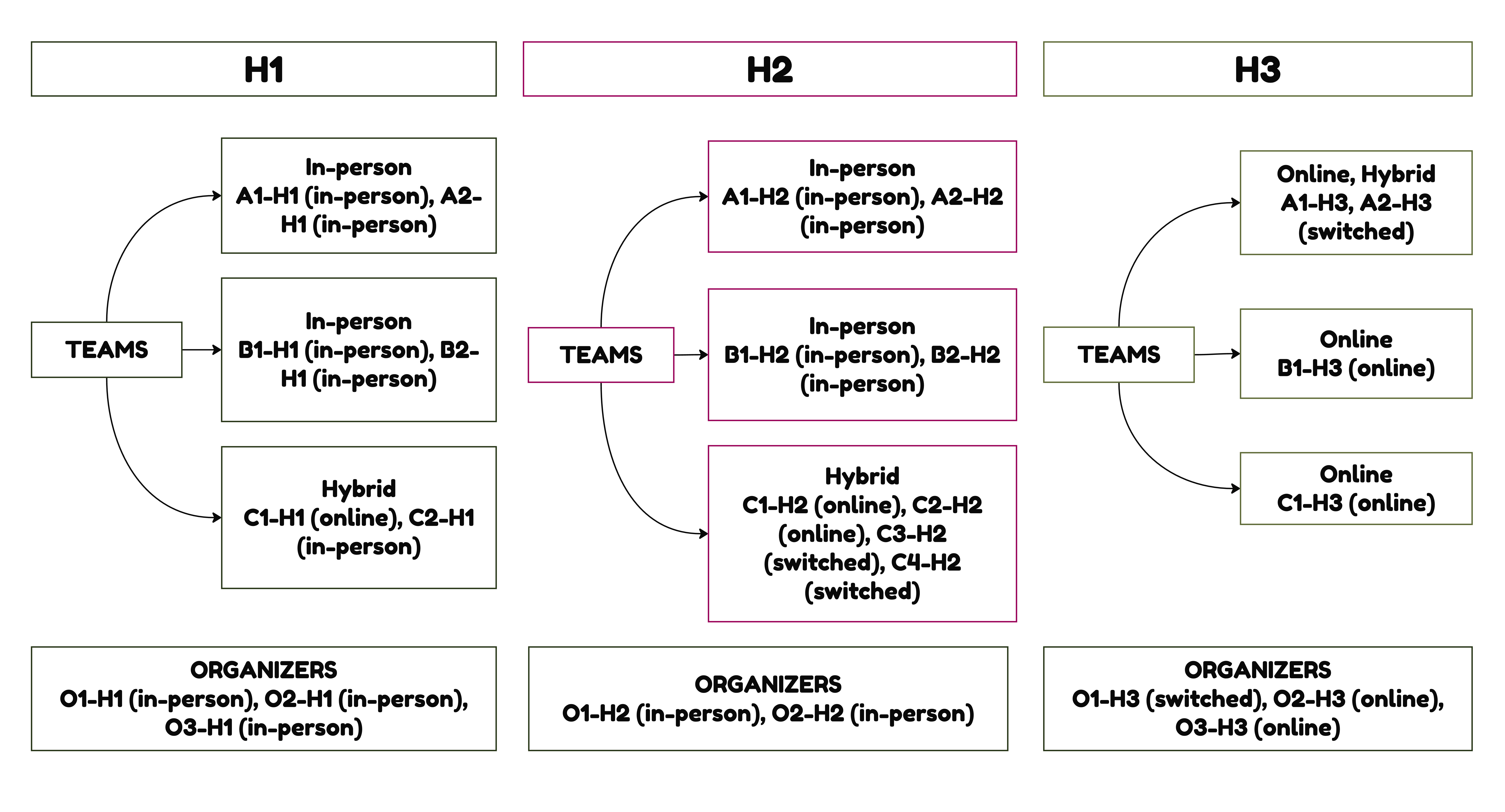} 
    \vspace{-20pt}
    \caption{Study participants — \textbf{hackathon participants and organizers} — across the studied hybrid hackathons (H1, H2, H3). \\
    \textbf{Teams} are identified by their overall mode of participation: \textit{in-person, online, or hybrid}, with \textbf{hybrid} teams including both in-person and remote participants at any point during the event.\\
    \textbf{Participants} are also labeled by their individual mode of participation: \textit{in-person, online, or switched}. \textbf{\textit{Switched}} refers to participants who changed their primary mode of participation during the event (e.g., from in-person to online). \\ 
    Each hackathon involved \textbf{a distinct set of study participants}, with no observed overlap. %\\ 
}\label{Fig:studyparticipants}
 %\vspace{-10pt}
\end{figure}

\paragraph{Participants} 
To answer \hyperref[RQ2]{\hr{RQ2}}, we focused on the participant experience. We asked participants about their collaboration experience, the challenges they faced working in hybrid teams, and their overall satisfaction with the event’s structure. The aim was to gather in-depth qualitative data on participants' perceptions of the impact of the hackathon scaffolding on their teamwork and collaborative process.
This also allowed us to compare and contrast the unique challenges and insights between teams, directly informing our understanding of collaboration at hybrid hackathons.
From the participants who showed interest, we selected 3 teams per hackathon (c.f. Fig.\ref{Fig:studyparticipants}) based on their different team configurations (hybrid, in-person, and online teams) to capture a broad spectrum of experiences in understanding the impacts of the hackathon scaffolding. 
We aimed to interview at least two members per team -- including individuals who participated in-person and online for hybrid teams -- to capture multiple perspectives and avoid relying on a single viewpoint. In total, we interviewed 18 participants from 9 teams across hackathons (3 teams for each hackathon, c.f. Fig.\ref{Fig:studyparticipants}) 
Moreover, we decided to interview participants individually to allow them to speak freely and potentially reveal differences in their perceptions about how they collaborated as a team. 

The 45-minute post-hackathon interview consisted of four main parts.
We started by first asking the participants about their motivations to participate in the hackathon event and their mode of participation (e.g. \textit{``Why did you decide to participate [in person, remote]?'', ``How did you participate in the hackathon? [in person, remote]?''}). 
Second, we went through the hackathon stages and asked them about how they decided with whom to work together as a team, which projects they worked on, how they approached their collaboration, and which tools they used (e.g. \textit{``Did you join as part of a pre-formed group or decide to form a group at the event?'', ``Can you describe how you communicated with your groupmates?'', ``Was everyone on the same page?'', ``What technologies did you use as a team?''}).
Third, we asked participants about their perception of their hackathon artifact and contributions and any impact the hybrid format can have on the team outcomes (e.g.,\textit{ ``How did you put everything together?'', ``Did being [in person, remote] affect your ability to contribute? How so?'', ``Are you satisfied with the outcome of your project?''}).
Lastly, we asked participants about their perception of the event design and their perception of how it supported their collaboration as a team (e.g. \textit{``How did the event's organization affect your collaboration (e.g. mentoring checkpoints, games, food, etc.)?'', ``How well did you think your team collaborated? What was particularly good or bad about the experience?''}).  
These questions served as a basis to examine how the hackathon scaffolding facilitates or hinders collaboration and discuss the perceptions of team satisfaction and team outcome (artifact) due to the considered hybrid dimensions.

\subsection{Analysis procedure}\label{Sec:analysisprocedure} 
We transcribed the audio recordings of all interviews using an online transcription software~\footnote{Otter.ai}, then proceeded with data coding. 
The coding process began with a deductive approach, using predefined dimensions from the literature -- \textit{synchronicity, physical distribution, team dynamics, dynamic transitions, and technological infrastructure} -- as a basis for the initial codes. For example, in the analysis of organizer interviews, we coded organizer plans and strategies for managing synchronicity (e.g., handling synchronous and asynchronous communication among participants in their teams) and the challenges faced during implementation (\hyperref[RQ1]{\hr{RQ1}}). For participants, we examined how they experienced the organizers' scaffolding efforts around these dimensions and how they navigated teamwork within the hackathon scaffolding (\hyperref[RQ2]{\hr{RQ2}}).
Afterward, we applied the initial codes to each interview, with one researcher responsible for developing and assigning the codes and another researcher supporting by double-checking the appropriateness of each coding assignment (i.e., ensuring the code fit the given statement). As we progressed, we revised the list of codes, incorporating inductive coding to capture new themes emerging directly from the data, particularly those unique to the hybrid format. For instance, the research team discussed whether to introduce a new code when interviews introduced additional information that did not align with the existing codes. This process led to the identification of new codes related to synchronicity (e.g., \textit{time zone differences}), dynamic transitions (e.g., \textit{transitions between physical spaces}), technology infrastructure (e.g., \textit{tools suggested by organizers}, \textit{tools utilized by participants}, \textit{technology use over time}), and team dynamics (e.g., \textit{team modality composition}, \textit{hybrid team management}, \textit{sub-grouping}, \textit{parallel tasks}, \textit{leadership}). We iterated this process until no further codes emerged.
Once the final list of codes was applied to all interviews, we clustered the codes into four overarching themes, focusing on categories closely aligned with our research questions and the hybrid context. 
%

%\section{Findings}\label{Sec:findings}

\section{Findings}\label{sec:findings-hybrid-dimension}
In this section, we outline our findings related to organizer considerations of the hybrid dimensions (\hyperref[RQ1]{\hr{RQ1}}) and participant experiences of these considerations in the hackathon scaffolding (\hyperref[RQ1]{\hr{RQ2}}). These are documented by each hybrid dimension namely synchronicity (Section~\ref{sec:synchronicity}), physical distribution (Section~\ref{sec:phy-distribution}), dynamic transitions  (Section~\ref{sec:dynamic-trans}), technological infrastructure (Section~\ref{sec:technology}), and team size and dynamic (Section~\ref{sec:team-dynamic}). In addition, we will elaborate on additional considerations within the hackathon scaffolding that impacted the participant (Section~\ref{sec:dimension-others}).

\subsection{Synchronicity} \label{sec:synchronicity}

Hackathon participants needed to coordinate across time and modalities to plan tasks, allocate responsibilities, resolve technical issues, and integrate contributions. These collaborative activities were strongly influenced by the means of communication they used. 
Across the three hackathons we studied, synchronous and asynchronous means of communication played a critical role in facilitating coordination, ensuring shared awareness of activities and expectations, and resolving conflicts. However, in H1 and H2, issues arose not just from the tools themselves, but from a misalignment in communication expectations and inconsistent visibility between in-person and online participants. These may have affected participants' ability to stay on task, engage fully in team and event activities, and adequately respond to changes in the hackathon agenda. 

Organizers in H1 ensured that \textbf{key sessions were streamed via Zoom}. These sessions included the opening and closing ceremonies, seminars, judging, and mentor check-ins. This was an organizer strategy to make these events accessible in real-time for both in-person and online participants. However, we observed \textbf{no explicit strategies to synchronize communication between modalities}, and no interactions from online participants during the streamed group sessions (\textbf{obs.}~\footnote{We use ``\textbf{obs.}'' to mark findings that are based on observation notes.}).
For team communication and coordination, \textbf{organizers introduced a centralized online platform} -- Discord -- as a strategy to provide event updates to participants and to simplify the logistics of handling and communicating key engagements across modalities. For instance, the organizers introduced virtual judging, which required participants to be \textit{``in [Discord] channels [...] waiting to be judged''} (O1-H1).
While this strategy may have helped with some aspects of synchronization, issues emerged when \textbf{in-person participants overlooked or ignored online communications} because they were immersed in the co-located setting. Organizers expected participants to monitor Discord while attending in-person events. However, when these expectations failed, organizers adopted an in-situ workaround by issuing dual announcements -- where verbal announcements complemented Discord updates -- as an ad-hoc tactic. O3-H1 noted: \textit{``people in person just kind of tune it out [of Discord], [...] we have to make two announcements in two ways''} (O3-H1).
Despite these efforts, \textbf{online participants encountered challenges in gaining visibility and clarity on the event agenda}. As one organizer observed, \textit{``sometimes the online participants were confused as to what to do''} (O2-H1). 
This may have been a result of online participants of hybrid teams needing to actively negotiate synchronicity -- deciding when to be online, what needed live alignment, and how to maintain shared understanding, particularly in time-sensitive contexts like project judging. 
Teams in H1 also used synchronous check-ins to realign: \textit{``Everyone in our project was aware of what everyone else was doing''} (C1-H1). \textbf{Team members maintained situational awareness through quick updates in Discord or Zoom}, sometimes prompted by progress announcements rather than formal status checks. C1-H1 reported: \textit{``Like if someone finished that module of the project, and they would just say I'm finished''} (C1-H1).
However, maintaining \textbf{real-time awareness across modalities was uneven}. In-person participants could glance at each other's screens or tap teammates directly, while online participants sometimes experienced asynchronous lags in responses. For example, C1-H1 noted that showing work in person was easier than waiting for remote teammates to notice Discord messages:  \textit{``It was easier to just show someone in person what I was doing [...] but for the online guys, we had to wait until they saw the message or the shared screen, which slowed things down''} (C1-H1). C2-H1 found Discord’s screen sharing as a helpful tactic to bridge the lags, allowing both remote and in-person collaborators to view their screens:
\textit{``on Discord you can share your screen [...] so they could see actively what I was doing''} (C2-H1).
C2-H1 further preferred Discord's method of screen sharing, where multiple users can share their screen independently, and switch between others' screens as they wish:
\textit{``I also prefer this over in person [...] it's very annoying to try to see one person’s screen when others are also trying to look. [...] It’s very difficult to collaborate [...] without a full screen-share in front of your face.''} (C2-H1).
Sustaining synchronicity between participants in H1 also appeared to require \textbf{deliberate check-ins and additional coordination} as a tactic to overcome lags. As A1-H1 put it, \textit{``there was a little bit of a disconnect [...] because we weren’t available to answer each other''} (A1-H1). Additionally, participants introduced personal Zoom accounts to complement Discord and improve live collaboration—coordinating times asynchronously before switching to Zoom: \textit{`we had Discord [...] so we were texting each other about when we should have a Zoom call''} (A2-H1).

Participants in H2 relied on both synchronous and asynchronous communication to organize their work, seek support, and coordinate with others. Synchronous interactions were central to maintaining real-time alignment within and across teams. \textbf{Organizers utilized a persistent Zoom setup} as a strategy: \textit{``We decided that it’s simpler just to have one Zoom and multiple rooms so that people can connect to these rooms''} (O1-H2). This design aimed to support a continuous connection to the main event while enabling \textbf{private team discussions in in-person and virtual (Zoom) breakout rooms}.
Despite this setup, organizers recognized the difficulty of achieving an engagement balance across modalities: \textit{``Some people maybe don’t feel like they are gonna be heard. Like they’re speaking virtually into a conference room where everybody’s just kind of chit-chatting''} (O2-H2). This highlighted \textbf{social barriers to synchronous participation}, even when technical access was available, requiring additional facilitation to support balance across modalities. 
Participants in H2 also had to time and initiate their synchronous interactions with organizers. While floating mentors like O1-H2 and O2-H2 were available, O2-H2 explained that \textbf{support relied on participant initiative}: \textit{``It’s always communicated that they can reach out at any time [...] like come and ask questions -- we’re here to help''} (O2-H2). While this may have worked well for in-person participants or those comfortable with initiating their own synchronous interactions, \textbf{this may have placed an additional burden} on less confident individuals or on remote participants to time their requests appropriately. C4-H2, an online participant, noted challenges: \textit{``The communication was okay, but it could've been more clear [...] sometimes I wasn't sure where to ask''} (C4-H2). 
Discord was intended as a \textbf{shared asynchronous communication space}, but not all participants used it consistently, due to communication confusion or device limitations. C2-H2 noted: \textit{``some of us [...] were working from laptops where Discord isn't allowed. So we would have to log into our phone or something to check Discord''} (C2-H2).
To manage this, the team adopted a \textbf{supplementary channel strategy}, using email alongside Discord: \textit{``There was a little bit of mixed mode. We did some emails back and forth [...] some people notifying on the [...] Discord channel, but it was a little confusing''} (C2-H2).
\textbf{Inconsistent use of a dedicated channel may have caused lags} in responsiveness and planning, with some members missing key coordination moments. As C2-H2 reported: \textit{``Two out of our six members [...] missed [sync meetings]''} (C2-H2).

H3 demonstrated more intentional coordination of synchronous and asynchronous communication compared to H1 and H2. 
This intentionality was reflected in multiple ways. 
Organizers implemented a strategy where \textbf{all participants, including those in person, would collaborate via planned collaboration sessions via Zoom using their laptops} to maintain parity in engagement. As O3-H3 explained, this setup was to avoid the exclusion of remote participants: \textit{``We encourage everyone to join on their laptops with cameras on, even if they’re in the same room, to avoid excluding online participants''} (O3-H3). We observed that the use of \textbf{Zoom, as the shared synchronous space, was generally adhered to} during collaboration sessions, even when participants were co-located (\textbf{obs.}).
However, it is reasonable that offline interactions outside of the Zoom sessions could occur. A2-H3 noticed that for remote participants: \textit{``Sometimes the room is muted, but people are actually speaking together''} (A2-H3). A2-H3 also added that participants at \textit{``the hubs [were] not very interacting with online as much as you would have if they are all of them online''} (A2-H3).
\textbf{Asynchronous coordination appeared more integrated} in H3 than in H1 and H2. Organizers provisioned dedicated channels for ongoing discussions and announcements, project contributions, and collaborative documentation. A1-H3 described the contribution process: \textit{``We communicate on Slack during the event [...] I would always ping on Slack''} (A1-H3). They also described using Framapad to maintain awareness and documentation of task updates: \textit{``There’s a documentation [Framapad] where you [...] write about what you want to achieve [...] what others are working on''} (A1-H3). Organizers also supported GitHub-based workflows: \textit{``Most of the contributions happen on GitHub, so that’s pretty well defined''} (O3-H3).
Importantly, \textbf{participants adapted by utilizing specific tools to fit their individual needs}. B1-H3, a remote participant, explained that they worked after hours and relied on GitHub as their main interaction space, as opposed to Slack: \textit{``Most of the conversations were happening under the [Github] pull requests [...] I would check Slack once in a while, but mainly relied on GitHub and Zoom''} (B1-H3).
These measures reflect a more pre-structured and intentional scaffolding than seen in H1 and H2, where organizers offered fewer concrete participation guidelines and relied more on participant initiative.

\begin{table}[h!]
\centering
\resizebox{\textwidth}{!}{%
\begin{tabular}{|p{2.3cm}|p{4.3cm}|p{4.3cm}|p{4.3cm}|}
\hline
\textbf{Aspect} & \textbf{H1} & \textbf{H2} & \textbf{H3} \\ \hline

\textbf{Synchronous Coordination} & 
\begin{minipage}[t]{\linewidth}
\begin{itemize}[left=0pt,labelsep=0.5em]
    \item Key sessions streamed via Zoom (e.g., ceremonies, judging, mentoring)
    \item No explicit organizer strategies to synchronize across modalities
    \item Teams used check-ins and quick updates for situational awareness
\end{itemize}
\vspace{0.1pt}
\end{minipage}
&
\begin{minipage}[t]{\linewidth}
\begin{itemize}[left=0pt,labelsep=0.5em]
    \item Persistent Zoom setup with multiple rooms for continuous connection
    \item Private team discussions in-person and via virtual breakout rooms
\end{itemize}
\vspace{0.1pt}
\end{minipage}
&
\begin{minipage}[t]{\linewidth}
\begin{itemize}[left=0pt,labelsep=0.5em]
    \item Planned Zoom-based collaboration sessions, structured synchronous work
    \item All participants encouraged to join via their own laptops regardless of modality
    \item Zoom adhered to even by co-located participants
\end{itemize}
\vspace{0.1pt}
\end{minipage} \\ \hline

\textbf{Asynchronous Coordination} &
\begin{minipage}[t]{\linewidth}
\begin{itemize}[left=0pt,labelsep=0.5em]
    \item Discord used for updates, but adoption was uneven
    \item Teams supplemented with personal tools (Zoom, iMessage)
    \item Multi-stream screen sharing enabled remote–in-person collaboration
\end{itemize}
\vspace{0.1pt}
\end{minipage}

&
\begin{minipage}[t]{\linewidth}
\begin{itemize}[left=0pt,labelsep=0.5em]
    \item Discord available, but inconsistently used due to access/device issues
    \item Email used alongside Discord as fallback, asynchronous tactic
\end{itemize}
\vspace{0.1pt}
\end{minipage}
&
\begin{minipage}[t]{\linewidth}
\begin{itemize}[left=0pt,labelsep=0.5em] 
    \item Organizer-provisioned asynch tools (Slack, Framapad, and GitHub) used consistently by participants.
    \item Participants selected tools based on collaboration needs
\end{itemize}
\vspace{0.1pt}
\end{minipage} \\ \hline

\textbf{Challenges} &
\begin{minipage}[t]{\linewidth}
\begin{itemize}[left=0pt,labelsep=0.5em]
    \item Communication misalignment led to dual announcements
    \item Remote participants faced agenda visibility issues
\end{itemize}
\vspace{0.1pt}
\end{minipage}
&
\begin{minipage}[t]{\linewidth}
\begin{itemize}[left=0pt,labelsep=0.5em]
    \item Social barriers limited online engagement
    \item Uncertainty for remote participants around organizer access
    \item Mixed tool use (Discord, email) caused lags and planning issues
\end{itemize}
\vspace{0.1pt}
\end{minipage}
&
\begin{minipage}[t]{\linewidth}
\begin{itemize}[left=0pt,labelsep=0.5em]
    \item Offline interactions outside Zoom sometimes excluded remote participants
\end{itemize}
\vspace{0.1pt}
\end{minipage}\\ \hline

\textbf{Impact} & 
\begin{minipage}[t]{\linewidth}
\begin{itemize}[left=0pt,labelsep=0.5em]
    \item Shared awareness required effort to negotiate synchronicity
    \item Participant tactics (e.g., screen sharing, pings) aided awareness but coordination remained uneven
\end{itemize}
\vspace{0.1pt}
\end{minipage}
&
\begin{minipage}[t]{\linewidth}
\begin{itemize}[left=0pt,labelsep=0.5em]
    \item Engagement varied by initiative and access
    \item Async communication without structure led to coordination gaps
\end{itemize}
\vspace{0.1pt}
\end{minipage}
&
\begin{minipage}[t]{\linewidth}
\begin{itemize}[left=0pt,labelsep=0.5em]
    \item Integrated synchronous and asynchronous tools supported equitable collaboration
    \item Participants adapted tool use to fit personal collaborative needs within the infrastructure
\end{itemize}
\vspace{0.1pt}
\end{minipage} \\ \hline
\end{tabular}%
}
\caption{Summary of synchronicity findings across H1, H2, and H3}
\label{tab:synchronicity}
\vspace{-20pt}
\end{table}

\subsection{Physical Distribution} \label{sec:phy-distribution}
Across all hackathons, organizers recognized the importance of balancing physical and virtual spaces to support hybrid collaboration. However, the degree of integration between in-person and online participants varied.
%
%\subsubsection{Space Allocation and Use}

Our observation was that organizers designed H1’s \textbf{physical setup as a strategy to focus primarily on in-person participants and teams}, with virtual spaces such as Discord serving as supplementary platforms rather than being fully integrated into the event structure (\textbf{obs.}). This was likely because the organizers expected more in-person participation as opposed to a more balanced or significant online participation.
The physical distribution of space in H1 encouraged in-person socialization and informal networking, but at the cost of hybrid inclusivity. \textbf{Large communal areas} supported spontaneous interaction, as noted by B1-H1: \textit{``[...]We really made friends with the team next to us. And when we were on our break or something, we would just chat with them''} (B1-H1).
However, the \textbf{open-plan layout also led to noise and distractions}: \textit{``[...] more distractions happened in person because lots of noise and different people coming up to you and stuff like that''} (B1-H1). In response, B2-H1's team tactic was to relocate to quieter side rooms: \textit{``We went in a different room [...] not in the big room, but it was like a smaller room''} (B2-H1).
While these rooms helped reduce distraction, they were not dedicated team spaces nor optimized for hybrid engagement (\textbf{obs.}). 
In some rooms, organizers used 360-view video tools as a strategy to support mutual visibility and audio capture; however, \textbf{the side rooms were not dedicated to teams} but to smaller group events like seminars (\textbf{obs.}).
\textbf{Hybrid teams, therefore, had to independently manage their setups}, frequently alternating between physical and virtual spaces due to the lack of integrated hybrid infrastructure.
Participant C2-H1 highlighted how noise at the venue disrupted collaboration during key moments like pitch recordings, prompting a full shift to remote work as a tactic to mitigate these issues: \textit{``[...]The noise levels at the venue made it difficult to record our pitch video, so we transitioned to remote work''} (C2-H1). This highlights how H1’s physical distribution created barriers to hybrid collaboration and, in some cases, led teams to disengage from the in-person setting entirely (see Section~\ref{sec:dynamic-trans}).
Overall, while H1’s physical distribution facilitated in-person engagement, it overlooked the complexities of hybrid participation. \textbf{Without hybrid-friendly spaces, teams had to manage their own hybrid setups}, often leaving remote participants at the periphery, with their engagement setup limited to Zoom sessions and not embedded in the overall event structure. As one organizer acknowledged: \textit{``The online folks could join using Zoom [...] but we didn’t integrate them much beyond that''} (O2-H1).  

In H2, we observed that the organizers implemented strategies to deliberately integrate physical and virtual spaces for both in-person and hybrid teams (\textbf{obs.}). Unlike H1, where in-person collaboration was dominant, \textbf{H2 incorporated four (4) dedicated hybrid spaces equipped with Zoom setups} to facilitate communication between in-person and virtual participants.  This setup marked a beneficial improvement from H1, where such dedicated spaces were lacking. In H2, both in-person teams and in-person members of hybrid teams could also use these rooms, with Zoom connections to the communal room. 
C2-H2, who was initially in-person for the idea generation session of the hackathon, commented on the space setup: \textit{``The hybrid setup was also nice. The reason I was in [physical] breakout room three (3) is because it also had a Zoom connection [...] there were some people online that also are interested in the idea so I spoke with them with the Zoom setup''} (C2-H2).
Despite improvements, \textbf{noise remained a challenge, particularly in the communal spaces} where multiple teams worked in parallel. These areas enabled informal interactions and spontaneous networking, but also caused acoustic distractions, prompting teams to seek quieter environments. B1-H2 and B2-H2 described initially trying to find quieter corners before relocating to separate rooms as a tactic to improve focus: \textit{``Based on our experience on Friday, we know that the noise will probably disturb us, so we decided we need to go to a side room.''} (B2-H2). \textit{``There was a lot of parallel conversation going on in the main room. So we just moved out to the second floor [room] to avoid that, and have like a whiteboard of our own.''} (B1-H2).   
For virtual participants, \textbf{dedicated Zoom breakout rooms helped structure collaboration} by mimicking the spatial and task-based dynamics of in-person settings. Team C used them to divide into sub-teams that focus on specific tasks, and move between these virtual spaces as needed: 
\textit{``We had several breakout rooms, and people were just roaming around like in a physical room''} (C4-H2); \textit{``We split the work into three parts, and we asked for different breakout rooms [...] then we started communicating and working on what we needed to do''} (C4-H2).
Overall, H2 demonstrated notable improvements in hybrid space design over H1, particularly through the creation of dedicated hybrid-friendly rooms and the integration of Zoom in connecting physical spaces.

H3 adopted a different strategy by \textbf{prioritizing online engagement while treating physical hubs as supplementary spaces for local activities}.
Interviews with organizers suggest this orientation through their descriptions of the organization of physical venues -- limiting in-person access to just a few days, emphasizing the centrality of online tools, and framing physical hubs as optional, flexible, and supplementary. O1-H3 described the autonomy given to local hubs: \textit{``It’s up to the local hub organizer on how to get organized [...] as long as it aligns with the broader guidelines''} (O1-H3).
While hubs were equipped for hybrid collaboration, organizers deliberately limited their use to a few days rather than the full event to reduce logistical burdens, such as securing long-term space. As O3-H3 explained: \textit{``We do keep that for just a couple of days [...] It can be a lot to find a meeting room that's free for a whole week''} (O3-H3).
This arrangement may have supported more consistent online participation, as \textbf{most collaboration occurred in shared digital spaces rather than across physical and virtual settings}. H3 organizers enabled this virtual environment using hybrid tools like Slack, Zoom, and Framapad.
Participants in physical hubs could engage in person, but the focus remained online. C1-H3, who attended one of the hubs, described the setting as a shared workspace, where \textit{``[...] everybody was just working on their computers [...] then it was just like sharing a working space''} (C1-H3). Still, C1-H3 confirmed that virtual engagement remained central: \textit{``It was nice to have someone around in person for small doubts or a quick chat [...] but mostly, we were all working on our computers, [...] on Slack and Zoom''} (C1-H3). B1-H3, who participated fully online throughout the event, chose this mode as it fit their schedule and preferences: \textit{``I actually like the remote stuff, because in the end, I ended up doing it after working hours [...] so I could take a break and then contribute''} (B1-H3). 
The \textbf{strategy to treat physical spaces as secondary may have helped reduce noise challenges} seen in H1 and H2.  However, the flexibility afforded to local hubs meant that \textbf{participants' experiences may have varied depending on how those hubs were configured}, although we did not observe many differences (\textbf{obs.}). A1-H3, who participated remotely, noted their collaboration with in-person hub participants: \textit{``I got to see how they were working together and trying to collaborate with us [...] I saw that visually through the hub space''} (A1-H3).
Still, \textbf{maintaining a balance between online and in-person interaction remained a challenge}. A2-H3, the other team member of A1-H3 at a physical hub, \textbf{pointed out moments of disconnect when in-person participants focused on local conversations}: \textit{``Inside the hub, we have this huge screen with Zoom, but sometimes there's a disconnect [...] people are muted, chatting amongst themselves in person, and the online people don’t hear everything''} (A2-H3). 
We summarize our findings in Table~\ref{tab:synchronicity}.

\begin{table}[ht!]
\centering
\resizebox{\textwidth}{!}{ %
\begin{tabular}{|p{2.35cm}|p{4.3cm}|p{4.3cm}|p{4.3cm}|}
\hline
\textbf{Aspect} & \textbf{H1} & \textbf{H2} & \textbf{H3} \\ \hline

\textbf{Hackathon Space Design} & 
\begin{minipage}[t]{\linewidth}
\begin{itemize}[left=0pt,labelsep=0.5em]
    \item Large communal areas and smaller side rooms provided
    \item No dedicated hybrid spaces; setup favored in-person teams
    \end{itemize}
\vspace{0.1pt}
\end{minipage}
& 
\begin{minipage}[t]{\linewidth}
\begin{itemize}[left=0pt,labelsep=0.5em]
    \item Main communal space plus hybrid rooms with Zoom setups
    \item Dedicated breakout rooms supported a hybrid team structure
    \end{itemize}
\vspace{0.1pt}
\end{minipage} 
& 
\begin{minipage}[t]{\linewidth}
\begin{itemize}[left=0pt,labelsep=0.5em]
    \item Local hubs as supplementary to a virtual-focused agenda
    \item Hubs open for short durations under shared guidelines
    \item Collaboration centered in shared digital spaces
    \end{itemize}
\vspace{0.1pt}
\end{minipage}
\\ \hline

\textbf{Noise and Disruptions} & 
\begin{minipage}[t]{\linewidth}
\begin{itemize}[left=0pt,labelsep=0.5em]
    \item Communal spaces were noisy, disrupting hybrid teams
    \item Smaller rooms were quieter but not hybrid-friendly
    \end{itemize}
\vspace{0.1pt}
\end{minipage} 
& 
\begin{minipage}[t]{\linewidth}
\begin{itemize}[left=0pt,labelsep=0.5em]
    \item Teams relocated from noisy spaces to quieter rooms
    \item Hybrid-dedicated rooms reduced disruptions
\end{itemize}
\vspace{0.1pt}
\end{minipage} 
& 
\begin{minipage}[t]{\linewidth}
\begin{itemize}[left=0pt,labelsep=0.5em]
    \item Limited use of hubs minimized noise-related issues
 \end{itemize}
\vspace{0.1pt}
\end{minipage}
\\ \hline

\textbf{Challenges} & 
\begin{minipage}[t]{\linewidth}
\begin{itemize}[left=0pt,labelsep=0.5em]
    \item Noise disrupted hybrid collaboration
    \item No hybrid setups; teams managed on their own
\end{itemize}
\vspace{0.1pt}
\end{minipage}    
    &  
    
\begin{minipage}[t]{\linewidth}
\begin{itemize}[left=0pt,labelsep=0.5em] 
    \item Noise in communal spaces disrupted collaboration
\end{itemize}
\vspace{0.1pt}
\end{minipage}
& 
\begin{minipage}[t]{\linewidth}
\begin{itemize}[left=0pt,labelsep=0.5em]
    \item Occasional disconnects between local and virtual participant interactions
    \item Hybrid experience varied by hub configuration
    \end{itemize}
\vspace{0.1pt}
\end{minipage} \\ \hline

\textbf{Impact} & 
\begin{minipage}[t]{\linewidth}
\begin{itemize}[left=0pt,labelsep=0.5em]
    \item Hybrid teams struggled to coordinate across modalities and manage setups independently 
     \end{itemize}
\vspace{0.1pt}
\end{minipage}
& 
\begin{minipage}[t]{\linewidth}
\begin{itemize}[left=0pt,labelsep=0.5em]
    \item Hybrid spaces strengthened integration between modalities
    \end{itemize}
\vspace{0.1pt}
\end{minipage}
& 
\begin{minipage}[t]{\linewidth}
\begin{itemize}[left=0pt,labelsep=0.5em]
 \item Virtual-first design and limited hubs fostered more balanced engagement, though varied by hub
    \end{itemize}
\vspace{0.1pt}
\end{minipage} \\ \hline

\end{tabular}%
 }
\caption{Summary of physical distribution findings for H1, H2, and H3}
\vspace{-20pt}
\label{tab:physical-distribution}
\end{table}

\subsection{Dynamic Transitions} \label{sec:dynamic-trans} 

Across the three hackathons, the concept of dynamic transitions -- where participants switch between in-person and online modalities -- was present but not always explicitly scaffolded.

In H1, the hybrid model was maintained to promote accessibility, especially for those unable to travel to the venue: \textit{``We've kept the hybrid model just to keep kind of accessibility to students who can't travel here''} (O2-H1). Organizers also allowed participants to move freely between physical and virtual modes.
While inclusive in intent, this \textbf{hybrid flexibility lacked structured support when participants chose to transition} for various reasons. Some \textbf{participants adopted transitions as a tactic to manage their working conditions}. For example, B1-H1 chose to switch to remote participation later in the event for personal comfort: \textit{``I wanted to work from the comfort of my home, so I switched to remote work later in the event''} (B1-H1). Similarly, C2-H1 described moving online due to distractions: \textit{``The noise levels at the venue made it difficult to record our pitch video, so we transitioned to remote work''} (C2-H1). 
These \textbf{participant-driven, informal transitions led to collaboration breakdowns}, as A2-H1 noted: \textit{``Online decision-making was a little bit messy after transitioning''} (A2-H1). 
However, some teams \textbf{navigated transitions by planning ahead, using in-person, synchronous periods for idea generation and task alignment, and leaving task execution for remote work}. These transitions also served as tactics to manage collaboration conditions. Team C, for example, used the opening hours on-site to coordinate: \textit{``I met with my team, there was one person in person with me, and then two virtual, and we all basically just called in, went over each prompt and decided which one we would want to target''} (C2-H1). During this early window, they made time-sensitive decisions -- such as selecting an interface design -- while still co-located. 
The team continued working asynchronously after transitioning, as C2-H1 described: \textit{``I left [...] to record my part of the video back in my dorm because it was very loud at the room that they were hosting it in [...]''} (C2-H1).

H2 organizers introduced slightly more deliberate strategies to support transitions. For instance, the presence of \textbf{a designated online organizer appeared to be an intentional strategy to provide a point of contact for virtual participants} (\textbf{obs.}), including those transitioning from in-person to online participation.
Still, \textbf{teams had to coordinate transitions independently}.
C2-H2, took advantage of the hybrid format to attend in person on the first day for networking and team formation: \textit{``I was there in person the first day to see who are the other participants and who might be interested in my idea [...] So it's good to meet with them in person''} (C2-H2). By the second day, they moved online to align with their remote team: \textit{``If I was on-site, I would be the only person on-site and then everybody else would be virtual on my team. So [...] I said it would be able to make more sense for me to be virtual as well so I can coordinate with them all remotely''} (C2-H2).   
Yet, as in H1, participants in H2 reported \textbf{coordination issues after transitions}. C3-H2 reflected on the slower pace of online work: \textit{``In hindsight, I believe that working in person would have allowed me to receive feedback more quickly or obtain necessary advice when my work wasn't proceeding smoothly''} (C3-H2). C2-H2 also commented that, \textit{``Two out of our six members were a little less engaged. So, for example, if we said we were gonna meet at two, they missed that''} (C2-H2).

H3 organizers' strategy of \textbf{limiting physical hub operation introduced structure to transitions and reduced unplanned shifts}. The event was designed so that in-person hubs operated only on selected days, while most of the event was conducted online (see Section~\ref{tab:physical-distribution}). 
Organizer O3-H3 noted: \textit{``We encourage people to keep in-person hubs to a day or two [...] to give people the opportunity to come into the online spaces and to participate there as well, so that we don’t wind up losing that broader connection to the global community''} (O3-H3). 
C1-H3 experienced a mix of participation modes, attending a physical hub briefly before continuing online: \textit{``I just booked two days for the week. That's like, I just attended like one day in-person''} (C1-H3). 
Following C1-H3, this setup may have \textbf{reduced the frequency of transitions by encouraging participants to plan their attendance in advance}.
However, \textbf{H3's strategy did not eliminate dynamic transitions} due to personal reasons such as illnesses: \textit{``On the first day, I was online, I was a bit sick [...] then on the second day, I joined in person. So it was a good experience to be able to work from home and then later join in person''} (A2-H3).   
Still, it is possible that \textbf{reduced unplanned transitions may have led to a more stable collaborative environment}, contrasting the more reactive and participant-driven transitions observed in H1 and H2.
We summarize our findings in Table~\ref{tab:dynamic-transitions}.

\begin{table}[h!]
\centering
\resizebox{\textwidth}{!}{%
\begin{tabular}{|p{2.5cm}|p{4cm}|p{4cm}|p{4cm}|}
\hline
\textbf{Aspect} & \textbf{H1} & \textbf{H2} & \textbf{H3} \\ \hline

\textbf{Scaffolding and Planning} & 
\begin{minipage}[t]{\linewidth}
\begin{itemize} [left=0pt,labelsep=0.5em]
    \item  No structured plans or scaffolding for modality transitions
    \item  Teams coordinated transitions independently
\end{itemize}
\vspace{0.1pt}
\end{minipage}
& 
    \begin{minipage}[t]{\linewidth}
\begin{itemize}[left=0pt,labelsep=0.5em]
    \item Informal support via a designated online organizer
    \item Teams still coordinated transitions on their own
    \end{itemize}
\vspace{0.1pt}
\end{minipage}
& 
    \begin{minipage}[t]{\linewidth}
\begin{itemize}[left=0pt,labelsep=0.5em]
    \item Hub availability planned for limited days
    \item Strategy may have limited physical attendance and reduced unplanned shifts
    \end{itemize}
\vspace{0.1pt}
\end{minipage}
\\ \hline

\textbf{Flexibility for Participants} & 
\begin{minipage}[t]{\linewidth}
\begin{itemize}[left=0pt,labelsep=0.5em]
    \item Participants occasionally switched between modalities
    \item Transitions were unstructured and based on personal circumstances (e.g., fatigue, noise)
    \end{itemize}
\vspace{0.1pt}
\end{minipage}  
&
    \begin{minipage}[t]{\linewidth}
\begin{itemize}[left=0pt,labelsep=0.5em]
    \item Flexibility valued, especially for aligning with remote teammates
    \item Post-transition collaboration sometimes lagged in responsiveness 
    \end{itemize}
\vspace{0.1pt}
\end{minipage}  
& 
    \begin{minipage}[t]{\linewidth}
\begin{itemize}[left=0pt,labelsep=0.5em]
    \item Participants had the option to attend physical hubs when available
    \item Virtual-first design reduced the need for frequent transitions
    \end{itemize}
\vspace{0.1pt}
\end{minipage}  
\\ \hline

\textbf{Challenges} & 
 \begin{minipage}[t]{\linewidth}
\begin{itemize}[left=0pt,labelsep=0.5em] 
    \item  Reactive transitions to remote work led to collaboration breakdowns
    \end{itemize}
\vspace{0.1pt}
\end{minipage}
&
     \begin{minipage}[t]{\linewidth}
\begin{itemize}[left=0pt,labelsep=0.5em]
    \item Post-transition coordination issues, delayed feedback, and affected cohesion 
\end{itemize}
\vspace{0.1pt}
\end{minipage}
& 
 \begin{minipage}[t]{\linewidth}
\begin{itemize}[left=0pt,labelsep=0.5em]
    \item Dynamic transitions still occurred despite planning constraints
\end{itemize}
\vspace{0.1pt}
\end{minipage}
\\ \hline

\textbf{Impact} & 
\begin{minipage}[t]{\linewidth}
\begin{itemize}[left=0pt,labelsep=0.5em]
    \item Dynamic transitions enabled needs-based participation
    \item Planned transitions helped maintain collaboration momentum
    \end{itemize}
\vspace{0.1pt}
\end{minipage}
&
\begin{minipage}[t]{\linewidth}
\begin{itemize}[left=0pt,labelsep=0.5em]
    \item Communication and engagement slowed during remote phases
\end{itemize}
\vspace{0.1pt}
\end{minipage}& 
\begin{minipage}[t]{\linewidth}
\begin{itemize}[left=0pt,labelsep=0.5em]
    \item Physical hub limits reduced unplanned transitions
    \item Virtual focus contributed to a more stable collaboration environment
    \end{itemize}
\vspace{0.1pt}
\end{minipage}\\ \hline

\end{tabular}}
\caption{Summary of dynamic transitions findings across H1, H2, and H3}
\label{tab:dynamic-transitions}
\vspace{-15pt}
\end{table}

\subsection{Team Size and Dynamic} \label{sec:team-dynamic} 

Across H1, H2, and H3, organizers implemented differing strategies for team size and dynamics, which may have impacted how participants formed teams, how leadership roles emerged and were defined within teams, as well as task distribution and teamwork over the hackathon period.

In H1, organizers recommended teams of 2--4 members, with \textbf{team formation arranged by organizers via Discord channels and in-person team formation events}. O2-H1 described the Discord team formation strategy: \textit{``We have a team formation channel in the discord for people to introduce themselves and say: I'm looking for X number of teammates, or this is me, I'd like to join a team''} (O2-H1). O2-H1 also noted the in-person team formation strategy: \textit{``[...] we offered team formation [event] at the beginning [of the event] about 8 am.''} (O2-H1). 
Following this, O1-H1 observed students forming teams organically: \textit{``I walked into the [event] room [...] there was like a couple students in there, so I was like -- ``Are you guys all on the same team now?'' They were like -- ``Yeah, I guess we are now''.''} (O1-H1).
A Zoom connection to the physical room was provided as a strategy to support online participation. However, no online interaction or hybrid teams were observed during the team formation event (\textbf{obs.}).
\textbf{Some team activities were not completely synchronized with the digital infrastructure}, leading to fragmentation in how teams engaged with organizers and access to shared resources.
For instance, some teams formed in person never registered on Discord, excluding themselves from organizer-coordinated mentoring and judging workflows (see Section~\ref{sec:technology}): \textit{``Some of the people formed teams in person here [...] but then they never created their teams in the Discord. So we have no way to [...] record the team exists [...] they kind of got excluded''} (O3-H1). 
Once teams were formed, \textbf{leadership roles were not formally assigned}. According to O2-H1, leadership often defaulted to whoever set up the team in Discord:  \textit{``Whoever makes the team on Discord [...] that member automatically becomes like the lead. We don't ask for them [...] we assume that the teams are gonna figure that out''} (O2-H1). However, there were other patterns of leadership assignment by technical expertise. C1-H1, an online team member of a hybrid team, assumed a leadership role due to their technical expertise, as participant C2-H1 explained: \textit{``[C1-H1] has the most background in this. He knew what needed to be done and he was in charge of what was going on''} (C2-H1).
In H1, \textbf{task distribution appeared to be driven by technical skill or resource availability}, with participants quickly assuming roles based on capability. For example, C1-H1 and C2-H1 discussed task distribution in their team, where C1-H1 handled programming, while others supported data collection, UI design, and project presentation. For example, C2-H1 reported: \textit{``[team member] who was virtual was in charge of UI design [...] while [C1-H1] was coding and kept track of what needed to be done for that part. And then [other team member] and I basically helped out what needed to be done like intern work [...]''} (C2-H1).
\textbf{Task distribution also appeared to be connected to pairing or sub-teaming}. Sub-teaming was often influenced by modality and time zones. B1-H1 confirmed this: \textit{``We decided to split up so we can essentially like gain more learning opportunities [...] we had different time commitments''} (B1-H1). C1-H1 also described close collaboration with another online member in their hybrid team: \textit{``[...] their skill set required more direct collaboration with me since the user interfaces directly impacted how I program the software''} (C1-H1). 
Overall, \textbf{H1 organizers provided basic scaffolding for team formation, role emergence, and task distribution}. However, H1 lacked mechanisms to ensure teams were consistently integrated into the digital infrastructure, contributing to fragmented coordination and uneven access to support.

In H2, team formation was guided by a suggested team size of 2 -- 7 members.  \textbf{Team formation occurred on the first day during a hybrid event}, where participants pitched ideas and explored shared interests.
C2-H2 described how pitching led to team building: \textit{``I was in person the first day [...] when we were presenting our ideas, I was able to go to a breakout room and [...] people came to chat [...] about the idea and how we were going to implement it''} (C2-H2). 
\textbf{Participants who proposed ideas often assumed informal leadership roles} and helped shape the team's direction. C4-H2 described their decision to join C2-H2's team, recognizing C2-H2 as the leader: \textit{``Yeah, [C2-H2] explains the idea, and I just saw that her idea has potential, and I can help her achieve what she wants''} (C4-H2). \textbf{Leadership roles sometimes extended to coordination tasks}, as C3-H2 noted:\textit{``The team leader consolidated the outputs [...] guided the roles [...] took on all three roles of note-taker, time-keeper, and presenter''} (C3-H2).
However, \textbf{larger teams may have introduced challenges for coordination and sustained participation}. One participant noted difficulties keeping all members engaged:  
\textit{``Then we got two more members that were interested [...] we really weren’t sure what we can find for them to do... [they] didn’t even attend the last day’s presentation so we had no idea if they were able to finish the work they wanted to do or not [...]''} (C2-H2). Reflecting on this, C2-H2 added: \textit{`` maybe if the teams were a little smaller [...] we could have had some tighter engagement.''} (C2-H2). 
In terms of how work was managed, C4-H2 described how \textbf{the team informally split tasks based on interest and modality}: \textit{``We asked who would like to do this part and then we just split it''} (C4-H2). This led to the \textbf{formation of sub-teams likely grouped by modality and time zones}. 
\textbf{Time zone distribution was seen to play a role in shaping collaboration} patterns, as C1-H2 noted: \textit{``[C4-H2] was also in the Europe time zone, and [C2-H2] was in the time zone of the hackathon [Eastern Standard Time], so the time zone was a bit of a spread because of this. I was generally with [C4-H2]''} (C1-H2). 
Participants requested virtual breakout rooms to support parallel collaboration: \textit{``We split the work into three parts, and we asked for different breakout rooms [...] then we started communicating and working on what we needed to do''} (C4-H2). Teams also planned moments to reconvene and reintegrate their work: \textit{``We met in our breakout room [...] then we asked the organizers to create additional breakout rooms for us''} and \textit{``Within the hour, we were going to meet back in the bigger breakout room for our team''} (C2-H2). 
Compared to H1, \textbf{H2 showed greater scaffolding of team formation and support} but lacked follow-through structures to support role clarity and task distribution.

In H3, there were \textbf{no predefined team sizes}, as \textbf{team formation was reported to be self-directed} and reported during the application process. O3-H3 explained: \textit{``we [...] use the onboarding sessions to [...] do that initial match up with people, and then we do expect participants to find their own way [...]''} (O3-H3). O2-H3 further described this design as \textit{``a feature, not a bug''} (O2-H3), and O3-H3 added that this design aimed to support long-term engagement: \textit{``We allow that flexibility [...] because we want them to stay engaged for the longer term''} (O3-H3).  
\textbf{Participants were free to work independently or as part of a team} under this open structure,  with new collaborations sometimes emerging through shared interests. 
O3-H3 explained: \textit{``You can work on whatever you'd like to work on [...] we don't expect people to be too rigid and say -- This is the group you're working in for the rest of the week.''} (O3-H3). Participant A1-H3 confirmed this strategy: \textit{``You could work as a team and also work independently''} (A1-H3).
However, C1-H3 described \textbf{difficulty finding collaborators}: \textit{``[...] I didn't get anyone else involved [...] I think it [the topic] wasn't just interesting''} (C1-H3). Some collaborations were also short-lived or fluid, as B1-H3 noted: \textit{``Sometimes people just collaborated on something for a little bit and then branched off again''} (B1-H3).
This strategy also influenced how roles were handled and tasks distributed. In H3, \textbf{roles were not formally defined, and teams were responsible for independently coordinating their contributions}. Contributions were provided incrementally, allowing for team members to make direct contributions to the final work: \textit{``The progress was really incremental [...] on the first day I could decide what I was going to work on, second day fleshing it out [...] third day was review''} (B1-H3).
Compared to H1 and H2, \textbf{H3 placed the least structural constraints on team size and dynamics, resulting in more fluid collaboration} -- but compared to H1 and H2, H3 shifted more responsibility for coordination onto participants. 
We summarize our findings in Table~\ref{tab:team-size-dynamics}.

\begin{table}[ht!]
\centering
\resizebox{\textwidth}{!}{%
\begin{tabular}{|p{2.1cm}|p{4.0cm}|p{4.1cm}|p{4.4cm}|}
\hline
\textbf{Aspect} & \textbf{H1} & \textbf{H2} & \textbf{H3} \\ \hline

\textbf{Team Formation} & 
\begin{minipage}[t]{\linewidth}
\begin{itemize}[left=0pt,labelsep=0.5em]
    \item  Teams of 2--4 recommended. 
    \item Formed via Discord and in-person events
    \item Some in-person teams were not fully integrated into the infrastructure 
    \end{itemize}
\vspace{0.1pt}
\end{minipage}    
    &
    \begin{minipage}[t]{\linewidth}
\begin{itemize}[left=0pt,labelsep=0.5em]
    \item Teams of 2--7 recommended. 
    \item Formed on the first day during a hybrid event
    \item Formed around shared interests
\end{itemize}
\vspace{0.1pt}
\end{minipage}    
&
\begin{minipage}[t]{\linewidth}
\begin{itemize}[left=0pt,labelsep=0.5em]
    \item No predefined team sizes; self-directed formation
    \item Participants were free to work independently or in teams
    \item Some collaborations were fluid or short-lived
\end{itemize}
\vspace{0.1pt}
\end{minipage}    
\\ \hline

\textbf{Role Emergence} & 
 \begin{minipage}[t]{\linewidth}
\begin{itemize}[left=0pt,labelsep=0.5em]
    \item Leadership informal; tied to technical expertise or team creation on Discord
    \end{itemize}
\vspace{0.1pt}
\end{minipage} 
    &
     \begin{minipage}[t]{\linewidth}
\begin{itemize}[left=0pt,labelsep=0.5em]
    \item  Idea proposers often assumed leadership
    \item  Leadership included coordination responsibilities
    \end{itemize}
\vspace{0.1pt}
\end{minipage} 
&
 \begin{minipage}[t]{\linewidth}
\begin{itemize}[left=0pt,labelsep=0.5em]
    \item Roles not formally defined   
    \end{itemize}
\vspace{0.1pt}
\end{minipage} 
\\ \hline

\textbf{Task Distribution} & 
\begin{minipage}[t]{\linewidth}
\begin{itemize}[left=0pt,labelsep=0.5em]
    \item Tasks likely assigned based on interest
    \item Sub-teaming by modality and time zone supported parallel work
    \end{itemize}
\vspace{0.1pt}
\end{minipage}
&
    \begin{minipage}[t]{\linewidth}
\begin{itemize}[left=0pt,labelsep=0.5em]
    \item Sub-teams formed by interest, modality, and time zone similarity 
   \end{itemize}
\vspace{0.1pt}
\end{minipage}
& 
    \begin{minipage}[t]{\linewidth}
\begin{itemize}[left=0pt,labelsep=0.5em]
    \item Teams and individuals coordinated contributions independently
    \item Participants could contribute incrementally to the final work
    \end{itemize}
\vspace{0.1pt}
\end{minipage}
\\ \hline

\textbf{Challenges} & 
\begin{minipage}[t]{\linewidth}
\begin{itemize}[left=0pt,labelsep=0.5em]
    \item In-person team formation not synchronized with digital systems
    \item Limited visibility for organizers or mentors to provide support
    \end{itemize}
\vspace{0.1pt}
\end{minipage}
& 

\begin{minipage}[t]{\linewidth}
\begin{itemize}[left=0pt,labelsep=0.5em]
    \item Managing large teams, complicated coordination, and sustained engagement 
    \end{itemize}
\vspace{0.1pt}
\end{minipage}
& 

\begin{minipage}[t]{\linewidth}
\begin{itemize}[left=0pt,labelsep=0.5em]
    \item Absence of fixed team structures made it difficult for some to initiate and sustain collaboration
    \end{itemize}
\vspace{0.1pt}
\end{minipage}
\\ \hline

\textbf{Impact} & 
\begin{minipage}[t]{\linewidth}
\begin{itemize}[left=0pt,labelsep=0.5em]
    \item Basic scaffolding for team formation, role clarity, and task distribution
    \item Limited mechanisms for consistent digital integration
    \end{itemize}
\vspace{0.1pt}
\end{minipage}
    &
    
\begin{minipage}[t]{\linewidth}
\begin{itemize}[left=0pt,labelsep=0.5em] 
    \item More structured scaffolding supported formation and cross-modality support
    \item Sub-teaming and time zone grouping shaped collaboration patterns
    \end{itemize}
\vspace{0.1pt}
\end{minipage}
& 

\begin{minipage}[t]{\linewidth}
\begin{itemize}[left=0pt,labelsep=0.5em]
    \item Basic scaffolding for team formation and support supported sustained engagement and individual autonomy
    \item Outcomes still relied on self-managed participation
    \end{itemize}
\vspace{0.1pt}
\end{minipage}
\\ \hline

\end{tabular}%
}
\caption{Summary of team size and dynamics findings across H1, H2, and H3}
\vspace{-20pt}
\label{tab:team-size-dynamics}
\end{table}

\subsection{Technological Infrastructure} \label{sec:technology}

Organizers in H1, H2, and H3 adopted different strategies for technological infrastructure to scaffold hybrid participation. Their choices in platforms, support mechanisms, and communication channels shaped participant engagement across modalities. We examined the organizer tool setup and intent, participant alignment and adaptations, and observed limitations and challenges across the three events.

Organizers in \textbf{H1 implemented a single central infrastructure --  Discord -- for communication, mentoring, and judging}, as a strategy to unify participation across in-person and online modalities. O1-H1 explained the rationale for the use of a central infrastructure even for judging: \textit{``We did all of the judging virtually [...] to make it easier for everybody. So we had all the judges in the Discord server [...] they had their own private judging room channels''} (O1-H1). While organizers introduced Zoom for large-group activities like workshops, they did not provide Zoom for team-based use. 
Despite the intent of streamlining interaction through a single tool, \textbf{the strategy with Discord assumed uniform adoption, which did not occur}. As a result, some participants and organizers improvised tactical workarounds. 
\textbf{Participants unfamiliar with Discord faced difficulties}: \textit{```We always have people [...] that are running into issues with Discord [...] sometimes people are really unfamiliar with the platform, so they need like a little extra help to get going''} (O3-H1 ). In-person participants also sometimes favored spontaneous, co-located communication, leading them to de-prioritize digital channels the organizer intended for use, as noted by O3-H1: \textit{``people in person just kind of tune it out [of Discord]''} (O3-H1).
Moreover, several in-person teams bypassed the platform entirely right after in-person team formation. As O3-H1 explained: \textit{``Some of the people formed teams in person here [...] but then they never created their teams in the Discord. So we have no way to [...] record the team exists [...] they kind of got excluded from that mentor check-in process''} (O3-H1).
These may have led participants to \textbf{turn to supplementary tools besides Discord as a tactic to improve their collaboration}. Some used their personal Zoom accounts for synchronous collaboration to supplement Discord: \textit{`[...] we had Discord communication [...] so we were texting each other about when we should have a Zoom call''} (A2-H1).
Others turned to messaging apps like iMessage, likely for informal or immediate updates:   
\textit{``We also had a iMessage group chat''} (B1-H1).
\textit{``we also used iMessage [...] but it was mostly Discord''} (C2-H1)
These participant-driven tactics suggest that, although organizers implemented a centralized infrastructure strategy around Discord, \textbf{participants in their teams gravitated toward tools and practices that better suited their communication styles and task progression}. This highlights a misalignment between the organizers' intended strategy and the participants' behavior.

In contrast, H2’s infrastructure was more loosely structured, with both \textbf{Discord and Zoom offered, but without an overarching integration strategy}. Zoom served as the default space for synchronous coordination, and Discord was available for asynchronous updates. O2-H2 explained Discord's role as the asynchronous layer: \textit{`` [...] we want Discord to be like a channel where people can just communicate help each other. And [...] if you have any problems at any level, you know, reach out to us [...]''} (O2-H2).
O2-H2 also emphasized Zoom's role:\textit{``we had the [Zoom] break rooms where they could be working in their own team [...]''} (O2-H2).
While organizers offered a dual-platform structure to support flexibility, \textbf{participants often adopted other tactics to improve accessibility, convenience, or workflow needs}. Discord posed challenges for those using restricted devices or multitasking, leading to missed updates. As C2-H2 noted: \textit{``We were working from laptops where Discord isn't allowed [...] we would have to log into our phone [...] when you're working [...] you may forget to check Discord''} (C2-H2).
O2-H2 also acknowledged a learning curve when dealing with Discord: \textit{``Discord is, it's not super difficult, but there's a learning curve to it''} (O2-H2).  
In response, participants adopted varying tactics to in response.
C2-H2's team adopted the use of email as a tactic to manage key updates, using it as a supplementary tool when Discord proved confusing: \textit{``We did some emails back and forth [...] Discord was a little confusing [...] so we would email each other''} (C2-H2). 
Participants also described using Google Docs to document contributions, track progress, and continue work beyond the event. C1-H2 reported: \textit{``We have like a big Google Docs. So we could see like what other people are doing''} (C1-H2); similarly, C2-H2 noted: \textit{``We created a Google Doc page [...] to capture our research for the hackathon and beyond''} (C2-H2). As in H1, these participant-driven tactics revealed gaps between the organizer's intent and the participants' behavior.
However, while tool redundancy wasn’t always cited as confusing, \textbf{simultaneous use of multiple platforms may have introduced coordination challenges} in some teams. C2-H2 mentioned: 
\textit{``There was a little bit of mixed mode [...] some people notifying on Discord, some by email [...] it was a little confusing''} (C2-H2). This again illustrates the gap between planned organizer intent in infrastructure planning and participant behavior in tool adoption, especially across hybrid teams that are navigating uneven access and usage norms.

H3 organizers implemented a \textbf{multi-tool strategy comprised of Slack, Github, Framapad, and Zoom} for hybrid teamwork and participation at the hackathon.
Zoom was used for real-time collaboration sessions, Slack facilitated team discussions and updates, GitHub served as the primary space for version control and contribution tracking, and Framapad enabled collaborative documentation and contributions across contribution sessions. Participants noted the usage of these tools: \textit{``Most of the conversations were happening under the [GitHub] pull requests [...] I would check Slack once in a while, but mainly relied on GitHub and Zoom''} (B1-H3); \textit{``We communicate on Slack during the event [...] I would always ping on Slack.''} (A1-H3); \textit{``There's a documentation [Framapad] where you get to also write about what you want to achieve during the event [...]''} (A1-H3) 
Unlike H1 and H2, participant reports in H3 suggest that they \textbf{generally engaged with the provided infrastructure as intended}. While we observed no mention of participants adopting tools outside of the official toolset, it is possible that some informal in-person communication occurred; however, this was not captured during interviews or in observation notes.
One potential reason for the higher alignment between organizer intentions and participant tool use in H3 may be in how \textbf{the organizers structured the tool ecosystem around distinct, purpose-driven functions}. Each tool played a clearly communicated role in the event workflow -- Slack for communication, GitHub for code collaboration, Framapad for documentation, and Zoom for synchronous sessions -- embedding usage into participants' work practices. This clarity likely reduced the need for improvisation and contributed to a more consistent experience across modalities.
\textbf{Accessibility was also prioritized in H3’s infrastructure design}, particularly given its online focus, which may have encouraged broader adherence to the intended infrastructure. O2-H3 discussed efforts to select tools compatible with screen readers and to document accessibility workarounds: \textit{``I started to get more involved in pressuring the community for more accessibility for screen reader users''} (O2-H3). O2-H3 further highlighted this need due to the accessibility issues in Slack: \textit{``Slack was very bad as far as accessibility, so we did a lot of documenting about recommendations for how to use Slack so that it’s accessible to people''} (O2-H3).
Still, even in H3, \textbf{the use of multiple platforms introduced some complexity}. O2-H3 acknowledged: \textit{``We need to streamline the technology [...] participants found it confusing to navigate between too many platforms''} (O2-H3).  O3-H3 also noted that some participants struggled with the number of communication links, suggesting that further streamlining of information could enhance usability:
\textit{``We've had feedback that sometimes we have stuff in too many different places, and people find it confusing. We're trying to simplify and streamline that''} (O3-H3). These reflections suggest that \textbf{while H3's infrastructure was more intentional and better adopted, it still needed some refinement} to reduce cognitive and logistical overhead.
We summarize our findings in Table~\ref{tab:hybrid-technology}.

\begin{table}[ht!]
\centering
\resizebox{\textwidth}{!}{%
\begin{tabular}{|p{2.3cm}|p{3.9cm}|p{3.9cm}|p{5.13cm}|}
\hline
\textbf{Aspect} & \textbf{H1} & \textbf{H2} & \textbf{H3} \\ \hline

\textbf{Organizer-Provided Tools} & 
\begin{minipage}[t]{\linewidth}
\begin{itemize}[left=0pt,labelsep=0.5em]
    \item   Discord as a central hub for communication, mentoring, and judging
\end{itemize}
\vspace{0.1pt}
\end{minipage} 
& 
\begin{minipage}[t]{\linewidth}
\begin{itemize}[left=0pt,labelsep=0.5em]
    \item  Discord for general communication
    \item  Zoom for hybrid meetings and virtual breakout rooms
    \end{itemize}
\vspace{0.1pt}
\end{minipage} 
&
\begin{minipage}[t]{\linewidth}
\begin{itemize}[left=0pt,labelsep=0.5em]
    \item  Slack for communication, GitHub and Framapad for collaboration, Zoom for synchronous meetings.
    \end{itemize}
\vspace{0.1pt}
\end{minipage} \\ \hline

\textbf{Participant behavior and adaptations} & 
\begin{minipage}[t]{\linewidth}
\begin{itemize}[left=0pt,labelsep=0.5em]
    \item In-person teams sometimes bypassed Discord, using verbal updates
    \item Some teams adopted personal Zoom or iMessage to support collaboration  
    \end{itemize}
\vspace{0.1pt}
\end{minipage} 
& \begin{minipage}[t]{\linewidth}
\begin{itemize}[left=0pt,labelsep=0.5em]
    \item Participants adapted tools to improve accessibility and workflow
    \item Email and Google Docs used for fallback communication when Discord access was limited
    \end{itemize}
\vspace{0.1pt}
\end{minipage} 
    &    
\begin{minipage}[t]{\linewidth}
\begin{itemize}[left=0pt,labelsep=0.5em]
    \item Tool ecosystem was function-driven and structured by organizers
    \item Participants largely followed intended tool usage and workflows
    \item Collaboration mostly occurred within this structured ecosystem
    \end{itemize}
\vspace{0.1pt}
\end{minipage} 
\\ \hline

\textbf{Challenges} & 
\begin{minipage}[t]{\linewidth}
\begin{itemize}[left=0pt,labelsep=0.5em]
    \item Discord adoption was uneven across modalities, limiting access to support
    \item Participants unfamiliar with Discord faced onboarding difficulties
    \end{itemize}
\vspace{0.1pt}
\end{minipage} 
& 
\begin{minipage}[t]{\linewidth}
\begin{itemize}[left=0pt,labelsep=0.5em]
    \item Discord's learning curve and device access issues created friction
    \item Use of multiple platforms introduced coordination challenges
    \end{itemize}
\vspace{0.1pt}
\end{minipage} 
& 
\begin{minipage}[t]{\linewidth}
\begin{itemize}[left=0pt,labelsep=0.5em]
    \item Navigating Slack, GitHub, Framapad, and Zoom created some complexity
    \item Participants may have found switching across tools overwhelming
    \end{itemize}
\vspace{0.1pt}
\end{minipage} \\ \hline

\textbf{Impact} & 
\begin{minipage}[t]{\linewidth}
\begin{itemize}[left=0pt,labelsep=0.5em]
    \item Misalignment between the organizer's intent and the participant's tool use
    \item Participants preferred tools that matched their communication and workflow needs
    \end{itemize}
\vspace{0.1pt}
\end{minipage} 
& 
\begin{minipage}[t]{\linewidth}
\begin{itemize}[left=0pt,labelsep=0.5em]
    \item Similar gaps between organizer expectations and participant behavior
    \item Supplementary tools created a ``mixed mode'', weakening synchronous/asynchronous alignment
    \end{itemize}
\vspace{0.1pt}
\end{minipage} 
& 
\begin{minipage}[t]{\linewidth}
\begin{itemize}[left=0pt,labelsep=0.5em]   
    \item No major gaps between the organizer's intent and the participant's behavior
    \item Accessibility-driven infrastructure supported more equitable engagement
    \item Despite some friction, the multi-tool strategy remained largely effective     \end{itemize}
\vspace{0.1pt}
\end{minipage}  \\ \hline

\end{tabular}%
}
\caption{Summary of findings for technological infrastructure in H1, H2, and H3}
\label{tab:hybrid-technology}
\vspace{-20pt}
\end{table}

\subsection{Additional hybrid considerations} \label{sec:dimension-others}

\subsubsection{Hybrid Mentoring and Organizer Support} \label{sec:hybrid-support-mentoring}
Across H1, H2, and H3, organizers implemented different strategies to manage hybrid mentoring and support for both in-person and remote participants. These strategies aimed to minimize discrepancies in how teams received mentorship and feedback.

In H1, organizers implemented a hybrid mentoring strategy using \textbf{Discord as the main platform for mentor assignment and engagement}. In-person and online mentors were required to check in with teams on Discord during specific time slots, regardless of the team modalities. O1-H1 emphasized the importance of these mentor check-ins: \textit{``We are doing that [mentor] check in mandatory [...] So from two to three and seven, eight, we are highly encouraging that students are in their Discord channel because they will be assigned, like one mentor to check in with [...] virtually hybrid, or both''} (O1-H1). 
This system was designed to promote mentor-student interaction, as many students were reluctant to ask for help. O2-H1 explained:\textit{``What we noticed was that the students were a little reluctant to go out and ask mentors, because it's just hard sometimes to ask for help. We're just [...] encouraging the interaction.''} (O2-H1). 
To reduce confusion, \textbf{H1 also opted for a fully virtual judging process}, ensuring that both in-person and remote participants engaged in judging in the same way. This decision aimed to create consistency, as reflected by O1-H1: \textit{``Last year, we did all virtual judging [...] we’re keeping the same model [this year] hoping to, you know, alleviate any confusion.''} (O1-H1). 
Participants used this hybrid mentoring check-ins and virtual judging, as A1-H1, part of an in-person team, noted: \textit{``[the team] met with the mentor in person the first time [...] and we did our second mentor check in over zoom and then we did all the judging over zoom.''} (A1-H1).
However, challenges arose with mentoring when \textbf{in-person teams who did not register on Discord were excluded} from important mentor check-ins, limiting their access to feedback and support. Additionally, the \textbf{virtual judging process faced logistical difficulties}. Some judges were not accounted for, creating confusion. O1-H1 reflected on this issue: \textit{``[...] definitely got a little bit hectic with the judging check-in, I think not everybody was accounted for. So we had whole companies kind of missing [...]''} (O1-H1). 
Participants also expressed \textbf{concerns about mentoring, with issues ranging from mentors not being able to find teams to mismatches in mentor expertise}. C2-H1 highlighted this problem:
\textit{``The first [mentor] couldn't find us at all [...] and the second one came over, and we were ready pretty far [...] I don't think he truly understood because our project was pretty advanced.''} (C2-H1).  

In H2, as a strategy, \textbf{organizers doubled as mentors}, providing both \textbf{an online and in-person mentor/organizer} to support remote and local participants. The dedicated online organizer supported remote teams, especially in the European time zone, while in-person organizers were accessible on-site or via Discord for those in the American (Eastern) time zone.
The \textbf{in-person organizers used a ``floating'' mentor strategy}, actively checking in with teams and encouraging them to reach out via Discord if more support is needed. As O1-H2 explained: \textit{``So we planned like just the floating type of mentorship [...] make the rounds, ask if [...] people have problems and [...] encourage them to reach out in cases where they get stuck''} (O1-H2). This approach may have allowed for deeper engagement with teams who actively sought help, allowing mentors to focus on specific challenges.
O1-H2 in H2 described how \textbf{in-person participants could ask questions directly, while remote participants reached out via Discord}: \textit{``I could walk around this room checking, and oftentimes people did have questions. So it was both in person and questions on Discord''} (O1-H2). O2-H2 added: \textit{``[...] I also had a couple of one-on-one conversations [...] people feel that they can email me and direct message me at any time''} (O2-H2).
This support was valued even in the hybrid setup. C2-H2, who participated in-person on the first day, described: \textit{``[Mentor O2-H2] stepped into the physical breakout room [with dedicated Zoom connection to online teammates] and we were able to get like ask him all of the questions [...]''} (C2-H2). C4-H2, who attended online, shared: \textit{``[...] I sent them an email or on Discord that they reply back [sic].''} (C4-H2).
However, \textbf{the floating mentoring strategy mostly benefited in-person participants and may not have supported online participants} as effectively.  C1-H2, an online participant, shared:  \textit{``I think the only person that actually helped me was [online organizer] [...] because I wasn't sure like how to get started.''} (C1-H2). 
\textbf{Organizer support aligned with participant time zone, in addition to modality, appeared especially beneficial}. C4-H2 also appreciated the online organizer being in a similar time zone: \textit{``I know [online organizer] also is in Europe, so his time was similar to mine [...] it was quite managed well [...]''} (C4-H1). However, time zone differences may have limited access to mentors with specific skill sets. We observed that \textbf{technical support remained largely tied to in-person organizers} in the American time zone due to differing mentor expertise \textbf{(obs.)}.

In H3, \textbf{organizers provided mentor support to participants}.
While team formation was participant-driven (see Section~\ref{sec:team-dynamic}), \textbf{organizers scaffolded onboarding support to teams in advance}. As O3-H3 explained: \textit{``We ask people to say in their application forms, what they want to be working on [...] We also identify if they have training needs [...] We provide the opportunity to get trained up on GitHub [...] And then we also hold onboarding sessions [...]''} (O3-H3). 
O3-H3 elaborated on these sessions: \textit{``So we do a lot of thinking about this [...] we hold onboarding sessions where we ask people to share some specific goals [...] then start trying to [...] pick out key themes [...] and make sure that we try and connect those people up throughout the week''} (O3-H3).
During the hackathon, organizers held \textbf{multiple online contribution sessions} each day of the hackathon \textbf{with dedicated organizers acting as mentors to support participants}. 
As O3-H3 explained: \textit{``[...] so [organizers] take the responsibility of leading sessions and will offer support and advice to people who need it during that session [...]''} (O3-H3). By \textbf{centering support in the online space}, this strategy \textbf{appeared to promote equitable access to organizers across modalities}. While organizers were available via the dedicated hackathon Slack channel, we observed that in H3, there were \textbf{opportunities to leverage its hackathon community for mentoring support} (\textbf{obs.}), although it is unclear if this was an organizer strategy. We observed participants ask for guidance through Slack channels during and outside of scheduled contribution sessions and receive help from community members not officially part of the event but available in the hackathon channel (\textbf{obs.}).
This was echoed by B1-H3, who commented on the unexpected support received: \textit{``I wanted to merge my pull request, but someone else ended up merging it for me because I was too busy''}; \textit{``People [...] coming out of nowhere and saving the day, more than willing to help and review stuff''} (B1-H3). 
This highlights the \textbf{benefits of fostering a strong hackathon community, where community-driven support can supplement formal mentoring}, especially during unscheduled times when organizers may not be available.
We summarize our findings in Table~\ref{tab:hybrid-mentoring}.

\begin{table}[ht!]
\centering
\resizebox{\textwidth}{!}{%
\begin{tabular}{|p{2.3cm}|p{4.1cm}|p{4.1cm}|p{4.9cm}|}
\hline
\textbf{Aspect} & \textbf{H1} & \textbf{H2} & \textbf{H3} \\ \hline

\textbf{Hybrid support structure} & 
\begin{minipage}[t]{\linewidth}
\begin{itemize}[left=0pt,labelsep=0.5em] 
    \item Mentoring and judging were fully virtual via Zoom and Discord
    \item Mentor check-ins scheduled through Discord 
\end{itemize}
\vspace{0.1pt}
\end{minipage} 
& 
\begin{minipage}[t]{\linewidth}
\begin{itemize}[left=0pt,labelsep=0.5em]
    \item Organizers served as mentors
    \item In-person floating mentors and a dedicated online organizer supported remote participants
    \item Online organizer covered a different (European) time zone 
    \end{itemize}
\vspace{0.1pt}
\end{minipage} 
&
\begin{minipage}[t]{\linewidth}
\begin{itemize}[left=0pt,labelsep=0.5em]
    \item Organizers acted as mentors
    \item Structured onboarding and multiple online contribution sessions; support via Slack
    \end{itemize}
\vspace{0.1pt}
\end{minipage} \\ \hline

\textbf{Participant engagement with hybrid support} & 
\begin{minipage}[t]{\linewidth}
\begin{itemize}[left=0pt,labelsep=0.5em]
    \item Scheduled mentor check-ins encouraged engagement
    \item Some participants negotiated ad-hoc in-person mentoring
    \end{itemize}
\vspace{0.1pt}
\end{minipage} 
& \begin{minipage}[t]{\linewidth}
\begin{itemize}[left=0pt,labelsep=0.5em]
    \item Online participants had to initiate contact for mentor support
    \item Time zone-aligned organizer support was especially helpful
    \end{itemize}
\vspace{0.1pt}
\end{minipage} 
    &    
\begin{minipage}[t]{\linewidth}
\begin{itemize}[left=0pt,labelsep=0.5em]
    \item Participants engaged through contribution sessions, Slack, and broader community mentoring support
    \end{itemize}
\vspace{0.1pt}
\end{minipage} 
\\ \hline

\textbf{Challenges} & 
\begin{minipage}[t]{\linewidth}
\begin{itemize}[left=0pt,labelsep=0.5em] 
    \item Some teams missed virtual mentoring/judging due to not registering on Discord
    \item Difficulty matching mentor expertise to team needs
    \item Virtual judging process had logistical issues
    \end{itemize}
\vspace{0.1pt}
\end{minipage} 
& 
\begin{minipage}[t]{\linewidth}
\begin{itemize}[left=0pt,labelsep=0.5em]
    \item Online participants received uneven support
    \item Time zone gaps limited access to mentors
    \item Technical support remained in-person-centric
    \end{itemize}
\vspace{0.1pt}
\end{minipage} 
& 
\begin{minipage}[t]{\linewidth}
\begin{itemize}[left=0pt,labelsep=0.5em]
    \item  Community-driven mentoring was unstructured and may not be sustainable
    \end{itemize}
\vspace{0.1pt}
\end{minipage} \\ \hline

\textbf{Impact} & 
\begin{minipage}[t]{\linewidth}
\begin{itemize}[left=0pt,labelsep=0.5em]
    \item Virtual mentoring and judging increased accessibility
    \item Support quality reduced by mentor mismatch and incomplete team integration into the infrastructure
    \end{itemize}
\vspace{0.1pt}
\end{minipage} 
& 
\begin{minipage}[t]{\linewidth}
\begin{itemize}[left=0pt,labelsep=0.5em]
    \item Support enabled deeper engagement for teams that actively reached out
    \item Despite some design considerations, mentoring remained uneven across time zones and modalities 
    \end{itemize}
\vspace{0.1pt}
\end{minipage} 
& 
\begin{minipage}[t]{\linewidth}
\begin{itemize}[left=0pt,labelsep=0.5em]   
    \item Structured support through contribution sessions and Slack ensured broad accessibility
    \item Online-centric support design enabled parity across modalities
    \item Community mentoring filled in when organizers were unavailable
     \end{itemize}
\vspace{0.1pt}
\end{minipage}  \\ \hline

\end{tabular}%
}
\caption{Summary of findings for hybrid mentoring and organizer support in H1, H2, and H3}
\label{tab:hybrid-mentoring}
\vspace{-20pt}
\end{table}

\subsubsection{Time zone coordination} \label{sec:timezone-coordination}
Across H1, H2, and H3, time zone coordination emerged as a critical factor for teams working in hybrid settings and organizers of open hybrid events. Time zone coordination impacted how tasks were distributed, how teams communicated asynchronously, and the effectiveness of collaboration.

In H1, \textbf{organizers focused primarily on the time zone of the physical venue} (East Coast), with limited consideration for remote participants in different time zones. 
O3-H1 admitted that \textbf{time zone differences were not fully incorporated into the planning process}, leading to a lack of real-time engagement for participants in distant time zones: \textit{``I don’t think we thought like too hard about, um, time differences and stuff''} (O3-H1).
As a solution, H1 organizers had the strategy of \textbf{recording key sessions like the opening and closing ceremonies} for later viewing. O2-H1 noted: \textit{``We do try recording all of our events [...] we just record them. So even if people couldn’t attend live, they just can go back and watch''} (O2-H1).

In H2, although the \textbf{organizers created the hackathon schedule and agenda around the American (Eastern) time zone}, \textbf{organizers were aware of participants from different time zones}, including European time zones and other American time zones. As a strategy, organizers planned the availability of a virtual organizer in the European time zone.
This supported the hybrid team in H2, which had participants in different time zones.  
Additionally, \textbf{the time zone differences allowed for shift-based collaboration among team members} to work asynchronously \textbf{and share progress across time zones}. This emerged as a tactic by the H2 team to maintain continuity despite time zone differences. C4-H2 described how their team worked in shifts to maintain productivity across different time zones: \textit{``We had different time zones, and maybe sometimes I work when other people in American time zone are sleeping and the other way around.''} (C4-H2).
C4-H2 explained the benefit of time zone differences:\textit{``[...] we are working [...] like 16 hours per day because people in Europe and the U.S. were working in shifts''} (C4-H2). 

In H3, the \textbf{hackathon schedule was designed around the European (UTC) time zone}, and \textbf{organizers took a more structured approach to time zone coordination} than in H1 and H2. Multiple collaboration sessions were offered daily -- from 08:00 to 21:30 UTC -- covering a range of time zones and enabling participation from regions including the Americas, where participants could join sessions at their suitable time.
O3-H3 explained: \textit{``We committed to supporting sessions that were run later in the day for people who were collaborating across the Americas time zones [...]''} (O3-H3). 
The combination of multiple daily sessions, extended event duration, and an online focus allowed participants to engage at times that best suited them, with organizers available across sessions to provide support.
Despite these efforts, O3-H3 acknowledged that \textbf{time zone coordination remained a challenge}, where participants may not have found a suitable time period in their time zone to join in the contribution sessions:
\textit{``It’s hard to find a solution that works for everyone [...] when you're working across multiple different time zones.''} (O3-H3). Still, \textbf{H3 stood out for its time zone-aware planning of the contribution sessions} across multiple time zones, which supported broader engagement and access to support for participants.
We summarize our findings in Table~\ref{tab:timezone-coordination}.

\begin{table}[ht!]
\centering
\resizebox{\textwidth}{!}{%
\begin{tabular}{|p{2.1cm}|p{4.1cm}|p{4.1cm}|p{4.5cm}|}
\hline
\textbf{Aspect} & \textbf{H1} & \textbf{H2} & \textbf{H3} \\ \hline

\textbf{Time zone planning} & 
\begin{minipage}[t]{\linewidth}
\begin{itemize}[left=0pt,labelsep=0.5em] 
    \item  Organizer planning focused on local (U.S. East Coast) time zone
    \item  Time zone coordination left to participants
    \item Organizers were not aware of participant time zones
\end{itemize}
\vspace{0.1pt}
\end{minipage} 
& 
\begin{minipage}[t]{\linewidth}
\begin{itemize}[left=0pt,labelsep=0.5em]
    \item Schedule based on American (Eastern) time zone
    \item Organizers were aware of participant time zones, but it wasn’t central to planning
    \item Teams worked across time zones but self-coordinated
    \end{itemize}
\vspace{0.1pt}
\end{minipage} 
&
\begin{minipage}[t]{\linewidth}
\begin{itemize}[left=0pt,labelsep=0.5em]
    \item Schedule aligned with European (UTC) time zone
    \item Multiple daily collaboration sessions covered different time zones
    \item Time zone differences were central to planning contribution sessions
    \end{itemize}
\vspace{0.1pt}
\end{minipage} \\ \hline

\textbf{Effect on collaboration} & 
\begin{minipage}[t]{\linewidth}
\begin{itemize}[left=0pt,labelsep=0.5em]
    \item  Key events (e.g., opening and closing ceremonies) were recorded for later viewing   \end{itemize}
\vspace{0.1pt}
\end{minipage} 
& \begin{minipage}[t]{\linewidth}
\begin{itemize}[left=0pt,labelsep=0.5em]
    \item  Time zone differences enabled shift-based collaboration
    \item  Teams used asynchronous tools and divided tasks by availability
    \end{itemize}
\vspace{0.1pt}
\end{minipage} 
    &    
\begin{minipage}[t]{\linewidth}
\begin{itemize}[left=0pt,labelsep=0.5em]
    \item  Contribution sessions attended according to personal time zone fit
    \item Staggered contribution sessions enabled broader participation
    \end{itemize}
\vspace{0.1pt}
\end{minipage} 
\\ \hline

\textbf{Challenges} & 
\begin{minipage}[t]{\linewidth}
\begin{itemize}[left=0pt,labelsep=0.5em]  
    \item Limited real-time engagement for distant time zones
    \item Time zone differences not fully addressed in planning
    \end{itemize}
\vspace{0.1pt}
\end{minipage} 
& 
\begin{minipage}[t]{\linewidth}
\begin{itemize}[left=0pt,labelsep=0.5em]
    \item  Communication inconsistencies when teams negotiated handoffs across time zones
    \end{itemize}
\vspace{0.1pt}
\end{minipage} 
& 
\begin{minipage}[t]{\linewidth}
\begin{itemize}[left=0pt,labelsep=0.5em] 
    \item Multiple time zone sessions helped, but didn't fully eliminate scheduling gaps
    \end{itemize}
\vspace{0.1pt}
\end{minipage} \\ \hline

\textbf{Impact} & 
\begin{minipage}[t]{\linewidth}
\begin{itemize}[left=0pt,labelsep=0.5em]
    \item Remote participants in other time zones faced fragmented involvement and missed shared events
    \end{itemize}
\vspace{0.1pt}
\end{minipage} 
& 
\begin{minipage}[t]{\linewidth}
\begin{itemize}[left=0pt,labelsep=0.5em]
    \item Shift-based handovers supported continuity, but required participant initiative
    \end{itemize}
\vspace{0.1pt}
\end{minipage} 
& 
\begin{minipage}[t]{\linewidth}
\begin{itemize}[left=0pt,labelsep=0.5em]   
    \item  Time zone - aware planning and distributed sessions supported broader engagement and access to support across regions
     \end{itemize}
\vspace{0.1pt}
\end{minipage}  \\ \hline

\end{tabular}%
}
\caption{Summary of findings for time zone coordination in H1, H2, and H3}
\label{tab:timezone-coordination}
\vspace{-20pt}
\end{table}

 \label{sec:findings}

\section{Discussion} \label{sec:discussion}
Reflecting on our findings, we will discuss how hybrid hackathon organizers prepared for and ran their respective events and how the event scaffolding might have influenced participants' perceptions of the event. The aim of this comparison is threefold: (1) to understand how and why organizers decided to run their event as they did, (2) to identify suggestions for improving organizer hybrid hackathon design and participant experience, and (3) to contribute to our understanding of hybrid collaboration settings.

\subsection{Implications for Theory}

\subsubsection{Organizer considerations of the hybrid format}
First, we discuss organizer considerations of the hybrid format in how they ran their events. Organizers in our study understood the importance of the hybrid dimensions (cf. Section~\ref{sec:findings-hybrid-dimension}) and considered them in varying degrees in creating their hackathon design. For example, considering \textit{synchronicity and asynchronicity} when managing hackathon activities and interactions (Table~\ref{tab:synchronicity}), \textit{physical distribution} when integrating the virtual environment into in-person spaces or creating dedicated hybrid spaces (Table~\ref{tab:physical-distribution}) or \textit{dynamic transitions} when managing how and why participant move between online and offline spaces (Table~\ref{tab:dynamic-transitions}). However, our findings suggest that organizers in our study did not prioritize this in practice, thus overlooking the impact of the hybrid format on collaboration at the event, and, as our findings suggest, led to diminished collaboration (through communication disconnects, participant disengagements, missing access to support, etc.). 
It appears that the organizers in our study focused more on the logistical demands of the event, including hackathon goals, participation size, demographics, and physical space availability.
Prior work also reports similar logistical concerns when organizing hybrid hackathons~\cite{gama2023comfort,khan2021innovation}, where we have seen the hybrid format as a way to provide flexibility but not as a part of the hackathon's design. However, a few~\cite{jonsson2022digital,porras2021experiences} have proposed design choices that have more consideration for the hybrid format.

Our findings suggest that hybrid hackathon design is shaped not only by logistical concerns (e.g., event size, goals, audience) but also by implicit decisions on how to scaffold hybrid collaboration.  
While all three hackathons were planned as hybrid events, H1 and H2 prioritized one modality over the other. In H1, organizers focused primarily on in-person participation, with remote involvement treated as secondary -- likely influenced by the large number of local university students and the 24-hour duration of the event.
In contrast, H3 exemplified a more premeditated approach to hybrid work, adopting an online-first model. Organizers integrated tool use with collaboration roles (e.g., Slack for updates, Framapad for documentation, GitHub for contributions), mandated norms like joining Zoom on individual laptops, and scheduled sessions across time zones. These structured strategies appeared more intentional, differing from the more reactive or participant-led approaches seen in H1 and H2.
Our findings also indicate that logistical factors, such as the hackathon's contributive goal, global participant demographics, and longer collaboration duration, influenced this decision. This design reduced the logistical burden of managing hybrid spaces but limited in-person engagement, resulting in a predominantly virtual experience. These insights suggest that event goals, scale, and participant dynamics shape organizers’ decisions around prioritizing a primary modality within a hybrid format.

However, this raises the question of whether such events can truly be called ``hybrid'' if they heavily favor one mode over the other. In our findings, H2 showed more consideration in managing between in-person and remote participation, providing dedicated hybrid spaces, Zoom breakout rooms, and dedicated organizers per modality to facilitate collaboration. However, this may have been easier to achieve in a more minor, controlled event with only 12 participants. It’s unclear if this balance would hold in larger, faster-paced events. 

\subsubsection{Effect of hackathon scaffolding on participants}

From our findings, certain organizer decisions regarding the hybrid dimensions (cf. Section~\ref{sec:findings-hybrid-dimension}) -- \textit{synchronicity and asynchronicity}, \textit{physical distribution}, \textit{dynamic transitions}, \textit{technological infrastructure}, and \textit{team size and dynamics} -- had specific influences on the participants.

For the synchronicity dimension, we observed that organizers sometimes struggled to address communication asymmetries. The case of H1 illustrates the limits of achieving symmetry in synchronous interaction in hybrid settings, even when organizers explicitly implemented strategies aimed at providing equal support across modalities by centralizing communication on a unified platform -- Discord. While organizers implemented these strategies -- tools and practices -- to support synchronization (e.g., virtual judging and mentoring, Discord as the main communication platform), in-person participants in H1 often neglected digital tools, leading to communication breakdowns (Table~\ref{tab:synchronicity}).
These tensions reflect what Bj{\o}rn~\textit{et al.}~\cite{bjorn2024achieving} describe as the impossibility of achieving symmetrical synchronous interaction in hybrid settings, where remote and co-located participants inevitably experience unequal access to interactional resources such as shared awareness and responsiveness.
This also suggests that co-located participants may require intentional support to avoid what Duckert~\textit{et al.}~\cite{duckert2023collocated} call ``collocated distance'' -- the mistaken assumption that physical proximity guarantees effective communication. Assumptions about seamless communication in co-located settings can thus be misleading; both in-person and remote participants are vulnerable to disengagement and misalignment. %
Similar to prior studies of hybrid teams~\cite{porras2021experiences}, we also observed how in-person dynamics can marginalize remote members (as seen in H2). Rather than aiming for perfect symmetry, our findings suggest that organizers should design hybrid events with persistent asymmetries in mind. In H1, organizers used in-person reminders to supplement digital announcements; in H2, participants structured work asynchronously to address time-zone differences. These adaptations align with Bj{\o}rn~\textit{et al.}'s call to develop interaction strategies tailored to hybrid realities~\cite{bjorn2024achieving}.
Such examples show how participants responded to organizer scaffolding with improvised adaptations -- a pattern evident not only in synchronicity but across other hybrid dimensions in this study. While a detailed exploration of strategies vs. tactics is beyond this paper’s scope, this interplay offers valuable directions for future work.

For the physical distribution dimension, we found that the organizer's design of physical spaces had impacts on the quality of teamwork, which, at times, influenced the participants' choice of modality. In H1, poorly designed physical spaces, such as noisy communal spaces, pushed participants to adapt by opting for virtual modes (Table~\ref{tab:physical-distribution}). When physical spaces fail to support in-person collaboration effectively, participants develop their own tactics, naturally transitioning to virtual interaction, which diminishes the hybrid experience and shifts the event towards a primarily virtual format. This shift can fragment the team and slow down coordination, especially when the dynamic transition dimension (Table~\ref{tab:dynamic-transitions}) is not considered by the organizers. 
Quiet, well-equipped hybrid spaces can encourage seamless transitions without forcing participants into one mode or the other.
In contrast, well-designed physical spaces can encourage in-person participation, enhancing collaboration by creating opportunities for face-to-face interaction, as highlighted by research showing that participants value in-person experiences when physical spaces are supportive~\cite{gama2023comfort}. If physical spaces lack incentives or ``perks'', participants will often default to remote tools, as noted by Porras~\textit{et al.}~\cite{porras2021experiences}.
In H3, the virtual-first model deliberately limited the role of physical spaces to align with the hackathon's global and online-focused goals. However, when in-person participation was planned, organizers adapted physical spaces for hybrid work to ensure they supported collaboration (Table~\ref{tab:physical-distribution}).

 For the dynamic transitions dimension, in addition to space transitions being influenced by the organizers' design of the physical space, the nature of the tasks participants were engaged in (Table~\ref{tab:dynamic-transitions}) also influenced transitions. This was seen in H2, where participants moved to remote work once the team was formed and initial project plans were made. The convenience and flexibility that the hybrid format offers encourage participants to make the best decisions for their teamwork. This tactic aligns with findings from Gama~\textit{et al.}~\cite{gama2023comfort}, where participants opted for remote work for convenience but returned to in-person collaboration when real-time interaction was necessary. 
However, the spontaneity of participant-driven transitions slowed team coordination, especially when moving from in-person to remote work (Table~\ref{tab:dynamic-transitions}). Transitions between in-person and remote modes, as seen in our study, introduced new asymmetries~\cite{bjorn2024achieving} that required tactics such as additional articulation work to realign team coordination after transition.
Our findings suggest that when organizers intentionally scaffold transitions between modes, fragmented collaboration and uneven participation issues can be alleviated while allowing participants to choose the best modality for their workflow. Organizers scaffolding physical spaces to facilitate rather than hinder collaboration, in addition to encouraging clear protocols for when participants switch between in-person and remote work, can benefit teams who choose to transition. 

In the technological infrastructure dimension, we found that technology integration shaped participant experiences but often depended on how well tools aligned with participant preferences and familiarity.
While organizers aimed to integrate a unified infrastructure, participants’ experiences varied across events. In H1 and H2, participants often pivoted to more familiar tools when the provided infrastructure didn’t meet their needs (Table~\ref{tab:hybrid-technology}), revealing a disconnect between organizer expectations and participant preferences~\cite{medina2019does}. In these cases, organizer scaffolding—though adequate in intent—did not always align with participant practices. Participants adapted tactics~\cite{bodker2016happenstance} such as using familiar tools over official ones. Even when tools were meant to be used consistently across modalities (e.g., in H1), experiences still differed by physical context, echoing Bj{\o}rn~\textit{et al.}’s findings on the persistence of interactional asymmetries~\cite{bjorn2024achieving}.
However, H3 demonstrated that even well-integrated tools can overwhelm participants when too many platforms are involved. While H3 participants reported a more inclusive cross-modal experience, some found the toolset complex and burdensome. This reflects findings in prior work~\cite{porras2021experiences}, which note that while suitable tools for hybrid events (e.g., Zoom, Slack, Miro) exist. However, without clear operational models, organizers and participants struggle to use them consistently and effectively across activities. 
Overall, the diversity of tools highlights the need for organizers to offer flexibility, allowing participants to use familiar tools while ensuring engagement with organizer-provided platforms for updates, coordination, and support. For instance, in H1, teams used iMessage and Zoom alongside the intended Discord. However, tool use was often improvised and varied across events, teams, and work phases, making it hard to trace stable or co-evolving ecologies as described by Lyle~\textit{et al.}~\cite{lyle2020s}.
At the micro level, we also observed a form of temporary artifact ecology, where participants brought their own tools into the event, overlapping or clashing with the organizer-provided tools. These overlaps often required negotiation over which ecology would guide collaboration. While we did not examine these dynamics in depth, future work could explore how such temporary ecologies rapidly form and evolve in short-lived collaborative settings.

For the team size and dynamics dimension, our findings suggest that the way teams are structured and roles are assigned has a significant impact on participants' ability to engage and coordinate across modalities. Organizer scaffolding, especially in larger teams, can unintentionally create modality-based sub-teams that isolate remote participants and disrupt intended hybrid collaboration (Table~\ref{tab:team-size-dynamics}). However, the influence of this scaffolding on team dynamics is limited; while organizers can shape initial team size, role guidance, and task distribution, team dynamics and leadership remain emergent, depending on participant interaction and familiarity.
Organizers’ efforts, such as early team formation opportunities and clear instructions on team size and composition, can help level the playing field for both in-person and remote participants. As our findings show, when roles were defined early, remote participants felt more engaged and could contribute effectively without being sidelined. For example, H1's recommendation for smaller teams lacked robust role structures, leading to fragmented collaboration as online and in-person members formed separate sub-teams. H2 also faced coordination challenges with larger teams and struggled with task distribution across modalities. H3, with a flexible, self-organized approach, allowed participants to move between individual and team tasks but relied on participant proactivity, which sometimes made it difficult to find collaborators.
These findings contribute to our understanding of hybrid teamwork, emphasizing the need for flexible yet structured scaffolding.

Our findings also indicated additional hybrid dimensions -- \textit{time zone coordination} and \textit{hybrid support and mentoring} -- within the hackathon scaffolding that shaped participant experiences of engagement and inclusivity. These dimensions, while often viewed as logistical details, played a role in enabling continuous collaboration and access to resources in ways that other hybrid dimensions (e.g., physical distribution, dynamic transitions) could not fully address.

For the time zone coordination dimension, we saw that limited attention to time zone coordination negatively affected participant engagement and task continuity across time zones. In contrast, H3’s structured coordination, with multiple collaboration sessions spread across time zones, allowed participants to work asynchronously and maintain productivity (c.f. Section~\ref{sec:dimension-others}). This approach allowed geographically dispersed participants to stay engaged without being disadvantaged by the event’s primary time zone, helping to prevent isolation and exclusion from real-time interactions.

Hybrid support and mentoring considerations are also vital in bridging the gap between in-person and remote participants, ensuring equitable access to guidance and resources. In H1, structured mentor check-ins offered both in-person and online support, but inconsistencies in mentor assignments limited their impact. H2 adopted a more balanced approach, with mentors and organizers available across modalities and time zones. H3 emphasized virtual mentoring, which participants viewed positively.
These cases highlight the value of intentional, modality-specific support, particularly for remote participants and those in different time zones.

\subsection{Implications for Practice}

\subsubsection{Organizer Suggestions}

Where intended, organizers should clearly communicate whether online or in-person participation will be prioritized, setting participant expectations early. 
To avoid overwhelming participants, organizers should consider minimizing fragmentation by limiting the number of tools introduced into the hackathon to a few core tools used to communicate with participants and manage the event. This ensures that important updates, mentoring access, and event-wide coordination are not missed. Additionally, simple onboarding tutorials are beneficial for getting participants familiar with the selected tools. 
While participants may choose to use additional tools within their own teams based on familiarity or preference, this parallel tool use should not be viewed as problematic -- as long as participants stay connected to the official communication channels.
Instead, organizers should ensure that both in-person and online participants receive timely and consistent information. Event-wide announcements should be posted in shared online channels. To help in-person participants stay attentive to these updates, organizers can reinforce important messages by briefly announcing or referencing them during physical interactions, reminding participants that detailed updates are posted in the communication channels.

Physical spaces should be flexible, offering quiet areas for calls and adaptable breakout rooms for meetings. If space is limited, a virtual-first approach can help reduce the reliance on physical spaces and enable smoother transitions between in-person and remote work. To avoid disruption during transitions, organizers should set up checkpoints where participants can decide their mode of engagement and use predefined transition points to prevent bottlenecks. Early team formation and leadership roles should be encouraged, and structured mentor support should include virtual rooms and office hours accessible to all participants.

\subsubsection{Participant Suggestions}

At the start of the hackathon, hybrid teams can fragment without clear leadership. Assign strategic roles for both in-person and remote participants to maintain coordination. Remote members can manage asynchronous tasks and check-ins, while in-person leaders handle live interactions. As collaboration progresses, regular check-ins keep everyone aligned. Sync points are especially important if subgroups form, as they help merge contributions.
During the hackathon, remote participants should maximize participation in synchronous sessions to stay visible and engaged. Document and share updates and next-step decisions on the team's shared platform, such as Slack or Discord, to ensure all members are on the same page. This is helpful before any periods of unavailability or if decisions are made while some members are absent. 
Hybrid hackathons offer flexibility between in-person and remote work, so use this to your advantage. Plan transitions if needed and set up regular check-ins with clear agendas to avoid delays.
Finally, stick with the tools your team agrees on. Avoid over-complicating things with too many platforms for communication and collaboration. Keep it simple to ensure everyone is on the same page.

\subsection{Limitations}
The goal of our study was to explore how organizers prepare for and run hybrid hackathons (\hr{RQ1}) and how the scaffolding of a hybrid hackathon can impact participants and their teams (\hr{RQ2}). It thus appeared reasonable to conduct an exploratory multiple case study~\cite{runeson2012case,wohlin2021case}. There are, however, limitations associated with this particular study design. 

Our study does not incorporate quantitative indicators, as our focus was not to assess productivity or output but to understand the collaborative dynamics within the hybrid setup -- something best captured through qualitative inquiry within the exploratory case study method. As such, we did not include metrics such as time-on-task or commit logs. While this limits generalization and performance-based comparisons across teams, it allowed us to gather in-depth insights into how participants and organizers navigated hybrid-specific challenges during the event.
That said, future research could extend this work by integrating qualitative insights with quantitative data (i.e., GitHub activity, performance metrics, or communication traces) to identify relations between organizer scaffolding strategies and participant collaboration and outcomes in hybrid hackathons.

We also studied a selected number of events that were organized by and attended by different individuals in different domains with different motivations and goals. While we made a deliberate selection of both events and teams across different dimensions (e.g., modality, size, communication practices), our aim was not to maximize coverage but to gain in-depth insights into collaboration within different hybrid hackathon contexts. We, therefore, perceive our sample to be sufficient for the aims of this exploratory study. Still, the insights gained are not meant to be generalizable beyond the contexts we studied. Rather, they offer rich insight into how collaboration evolved in different hybrid events.

While we highlighted the general location context of each hackathon (H1 and H2 on the campuses of large North American universities and H3 at a research institute in the United Kingdom), we acknowledge the difficulty of situating hybrid hackathons within a single cultural frame. These events are shaped by multiple influences -- including geographic location, institutional context, and the diverse backgrounds and intentions of organizers, mentors, and participants. As such, we treat each hackathon as a distinct case shaped by overlapping academic, professional, and community cultures rather than attempting to frame them as representative of a single national or institutional culture.

Moreover, the study was conducted by a team of researchers, which poses a potential threat to validity since different researchers observing and conducting interviews with different organizers and teams might lead to varying interpretations. To minimize this risk, we carefully planned the study by repeatedly discussing the design with the involved researchers. The presence of the researchers during the hackathons themselves can also affect the reported findings, despite our efforts to refrain from interfering during the events. We also abstained from making causal claims, instead providing a detailed description of observed behaviors and reported perceptions based on which we discuss differences in how organizers scaffold their events and how teams experienced them. We hope to see more research on how hybrid hackathons are organized and experienced to complement and extend our findings.

\begin{acks}
The contributions of Dr. Abasi-amefon Affia-Jomants and Dr. Alexander Nolte were partially funded by the Alfred P. Sloan Foundation under grant G-2023-20872. %number ACI-1547611. 
Special thanks to Reed Milewicz, Cal King, Meris Mandernach Longmeier, Audris Mockus, Anne Lee Steele, Esther Plomp, the hackathon mentors, participants, and the anonymous reviewers.

\end{acks}

\bibliographystyle{ACM-Reference-Format} %submission/template/
\bibliography{ref}

\end{document}